\documentclass[aps, prb, twocolumn, superscriptaddress, english]{revtex4-2}

\usepackage{graphicx}
\usepackage{cancel}
\usepackage{enumerate}
\usepackage{hyperref}
\hypersetup{pdfencoding=auto,colorlinks=true,linkcolor=blue,citecolor=blue}
\usepackage{textcomp}
\usepackage{amsmath}
\usepackage{amssymb}
\usepackage{bbold}
\usepackage{pgf}
\usepackage{soul}
\usepackage{subfigure}
\usepackage[english]{babel}
\usepackage{tikz}
\usepackage[normalem]{ulem}

\def\bra#1{\langle#1 |}
\def\ket#1{| #1\rangle}

\def\ud{\mathrm{d}}

\newcommand{\nep}{\textrm{e}}

\newcommand{\Ztwo}{{\mathbb Z}_2}

\graphicspath{{./}{./Figures/}}
\begin{document}
\title{Symmetry-protection Zeno phase transition in monitored lattice gauge theories }
\author{Matteo M. Wauters}
  \affiliation{Pitaevskii BEC Center, CNR-INO and Dipartimento di Fisica, Universit\`a di Trento,  Via Sommarive 14, I-38123 Trento, Italy}
 \affiliation{INFN-TIFPA, Trento Institute for Fundamental Physics and Applications, Via Sommarive 14, I-38123 Povo, Trento, Italy}
 
\author{Edoardo Ballini}
 \affiliation{Pitaevskii BEC Center, CNR-INO and Dipartimento di Fisica, Universit\`a di Trento,  Via Sommarive 14, I-38123 Trento, Italy}
 \affiliation{INFN-TIFPA, Trento Institute for Fundamental Physics and Applications, Via Sommarive 14, I-38123 Povo, Trento, Italy}
 
\author{Alberto Biella}
 \affiliation{Pitaevskii BEC Center, CNR-INO and Dipartimento di Fisica, Universit\`a di Trento,  Via Sommarive 14, I-38123 Trento, Italy}

\author{Philipp Hauke}
 \affiliation{Pitaevskii BEC Center, CNR-INO and Dipartimento di Fisica, Universit\`a di Trento,  Via Sommarive 14, I-38123 Trento, Italy}
 \affiliation{INFN-TIFPA, Trento Institute for Fundamental Physics and Applications, Via Sommarive 14, I-38123 Povo, Trento, Italy}

\begin{abstract}
   Quantum measurements profoundly influence system dynamics. They lead to complex nonequilibrium phenomena like the quantum Zeno effect, and they can be used for mitigating errors in quantum simulations. 
   Such an ability is particularly valuable for lattice gauge theories (LGTs), which require the challenging preservation of an extensive number of local conservation laws. 
   While it is known that tailored quantum measurements can soften violations of gauge symmetry, the nature of this protection, and in particular the possibility of a threshold behavior, is still unexplored. 
   Here, we demonstrate the existence of a sharp transition, triggered by the measurement rate, between a protected gauge-theory regime resistant to simulation errors and an irregular regime. 
   Our results are based on the paradigmatic example of a 1+1d $\Ztwo$ LGT. We study in detail the protection through projective measurements of ancillary qubits coupled to the local symmetry generators, and compare this approach with analog (weak) measurement protocols. 
   We show that, while the resulting ensemble averages in the continuous-time limit share the same Liouvillian dynamics, 
   different physical implementations of the stochastic gauge protection protocol yield trajectory unravelings with vastly different statistics. 
   Additionally, we design an on-chip feedback mechanism that corrects bit-flip errors and significantly enhances the discrete-time scheme. Our results shed light on the dissipative criticality of strongly-interacting, highly-constrained quantum systems, and they offer valuable insights into error mitigation and correction of gauge-theory quantum simulations. 
\end{abstract}

\maketitle
\section{Introduction}
Lattice gauge theories (LGTs) stand as key models at the crossroads of fundamental physics, condensed matter, and quantum information science~\cite{kogut1979,dalmonte2016, Preskill_arxiv2018, banuls2020}. 
Their unique framework provides a highly rewarding yet challenging target for quantum simulations \cite{Burrello2015, Martinez_Nature2016,Klco_PRA2018,Klco_PRD2020,Zohar_PhilTransA2021,Aidelsburger_LGT2021}, offering insights into the interplay between highly constrained dynamics~\cite{Celi_PRX2020,Surace_PRX2020,Indrajit_PRD2024}, possibly governed by nonabelian local symmetries \cite{Iqbal2024,Calajo_PRXQ2024}, intricate multi-body interactions, and the emergence of topological order with non-Abelian anyons~\cite{Iqbal2024,Satzinger_Science2021, Semeghini_Sci2021,Lumia_PRXQ2022,Shibo_ChinPhysLett2023,Andersen_Nature2023}. 
Quantum simulation protocols, however, are prone to imperfections and approximations that break the gauge symmetries, resulting in errors that lead to nonphysical results.
The local conserved quantities in LGTs present an intriguing avenue for harnessing these symmetries to implement several error-mitigation strategies~\cite{halimeh2022stabilizing,Cai_RMP2023}.
For instance, one can modify the Hamiltonian to include energy penalties on unphysical gauge sectors~\cite{Zohar_PRL2012, Banerjee_PRL2012, Kuno_NJP2015,Dutta_PRA2017,Halimeh_PRL2020,Halimeh_PRXQ2021, Halimeh_NJP2022, Mildenberger2025} design stochastic processes to destroy phase coherence between physical and nonphysical states~\cite{Stanningel_PRL2014, Lamm_arxiv2020, Kasper_PRD2023, ball2024zeno}, or use the gauge symmetry to improve the data quality in post-selection \cite{Mildenberger2025, Nguyen_PRXQ2022,cochran2024visualizing}. 

A particularly appealing approach is to adapt mid-circuit syndrome measurements, a cornerstone of quantum error-correction and detection algorithms~\cite{Devitt_RepProgPhys2013}, to the context of LGTs,  as the local gauge symmetries offer a natural qubit redundancy~\cite{Stryker_PRA2019, Rajput_npjQI2023, Raychowdhury_PRR2020,spagnoli2024faulttolerant}.
So far these proposals assumed fault-tolerant quantum computation or the possibility of performing mid-circuit measurements (together with immediate feedback) arbitrarily many times. 
The effect on the system dynamics of the interplay of error strength and the measurement frequency, also taking into account the further noise they might introduce, is still unexplored. 
A thorough understanding of the resulting stochastic dynamics can prove crucial for developing practical error-mitigation schemes for present-day and near-term quantum simulators, where significant extra resource requirements put full error-correcting capabilities still beyond reach.  

In this paper, we show that error protection in digital quantum simulations of LGTs emerges not gradually, but after a critical quantum Zeno transition. 
This threshold is determined by the ratio between the circuit density of local projective measurements~\cite{Wiseman_Milburn_2009, Sierant_PRB2022, Tirrito_SciPost2023} of the gauge symmetry generators and the coherent error strength. 
Framing our findings within the spectral theory of Liouvillians~\cite{Minganti2018}, we highlight the generality of the Zeno transition with respect to different measurement schemes in digital and analog protocols.
The presence of this transition has immediate consequences. For example, within the protected Zeno regime, our theory provides a systematic way to perform finite-error scaling to extrapolate exact expectation values. 
Importantly, the threshold appears already by only passively monitoring the gauge symmetry, which can be done via ancillas also as a delayed measurement, and which therefore is considerably simpler than active error correction. 
This measurement-induced protection also holds when taking into account weak depolarizing noise induced by the mid-circuit monitoring, where the Zeno phase survives for a finite range of measurement rates.
The combination of mid-circuit measurements with active feedback highlights the potential and limitations of rudimentary quantum error correction in quantum simulation protocols with both unitary and dissipative error channels.
Our results combine insights into stochastic criticality of monitored highly constrained systems with concrete consequences for state-of-the-art quantum simulation protocols. 

Our framework is based on the general theory of the interplay between monitoring and unitary dynamics, which introduces a further layer of complexity and can give rise to intricate many-body dynamics including, e.g., phase transitions in the entanglement properties~\cite{Cao2019,Alberton2021,Turkeshi2021,Piccitto_PRB2022, Tirrito_SciPost2023}. 
One remarkable phenomenon that emerges in this context is the celebrated quantum Zeno (QZ) effect, wherein frequent measurements hamper the quantum dynamics constraining the system evolution within a limited set of states~\cite{misra1977thezeno,peres80zeno,Itano_PRA1990,FACCHI_PLA2000,facchi2001from,facchi2002quantum,signoles2014confined,snizhko2020quantumzeno}.
Understanding and harnessing the QZ effect in a many-body system is currently one of the main challenges in the field of monitored systems and can be key for advancing the capabilities of quantum simulations and quantum information processing, as well as improving our comprehension of open quantum system dynamics \cite{syassen2008strong,patil2015measurement,froml2019fluctuation,froml2020ultracold,Biella_Quantum2021, Choi_PRL2021,Rosso_fermions_strong_2023,Rosso_bose_strong_trap_2022,Rosso_SUN_2022,maki2024loss}.

We focus on the effect of the measurement of local gauge symmetry generators, in close analogy to stabilizer measurements in the framework of error correction~\cite{Stryker_PRA2019,Rajput_npjQI2023,spagnoli2024faulttolerant} or state preparation~\cite{Tantivasadakarn_PRL2023, cobos2024noiseaware}, and the resulting quantum Zeno stochastic dynamics.
When these measurements are performed during the time evolution and not only at the end, we effectively realize a dynamical post-selection (DPS) protocol, allowing for both the detection and suppression of gauge violations in digital quantum simulations of LGTs.
We illustrate our findings mainly through the paradigmatic example of a $1+1$d $\Ztwo$ LGT with fermionic matter, where we explore the synergistic relationship between stochastic monitored quantum dynamics, error correcting codes \cite{Stryker_PRA2019,Satzinger_Science2021,Rajput_npjQI2023, Acharya_Nature2023, Bluvstein_Nature2024}, and strong constraints. We demonstrate the generality of our findings by analysing also gauge theories with $\mathbb{Z}_3$ and truncated $U(1)$ symmetry.

To identify the quantum Zeno transition, we study the Liovillian superoperator that generates the ensemble average of the DPS stochastic trajectories. Its spectrum allows us to distinguish between an error-protected phase and an ergodic one, separated by the quantum Zeno transition, determined by the ratio between the error strength and the measurement rate.
This suggests a reinterpretation of the proposal of Ref.~\cite{Stanningel_PRL2014}, taking the perspective where the gauge protection arises from continuous measurements of the gauge symmetry generators and not from their coupling with a noisy field.
Crucially though, despite both continuous and digital time evolution protocols relying on the same underlying Liuovillian dynamics, the measurement processes result in very different unravelings as stochastic quantum trajectories.

Previous works on using measurements to stabilize local symmetries have mostly focused on formal fault-tolerant frameworks~\cite{Lamm_arxiv2020,Kasper_PRD2023,Stryker_PRA2019,Rajput_npjQI2023,Raychowdhury_PRR2020, spagnoli2024faulttolerant}, often overlooking realistic quench dynamics and how measurements affect it, or on protecting against purely coherent gauge drifts~\cite{halimeh2022stabilizing,Halimeh_PRXQ2021,Halimeh_NJP2022,Mildenberger2025,Stanningel_PRL2014,Lamm_arxiv2020}, neglecting scenarios with both unitary and dissipative errors. 
Our scheme, instead, deals with both coherent and incoherent gauge-breaking errors within the same framework and, importantly, using the same circuit elements.
It efficiently compensates for a dissipative error channel, thanks to the natural redundancy of the gauge symmetries, while simultaneously enforcing a quantum Zeno protection against a coherent gauge drift.
Hence, the protected phase persists despite this considerably more complex scenario that combines both unitary and dissipative error sources. 
This finding is significant, as previous protocols protect against coherent errors or sufficiently colored noise, but not white noise \cite{kumar2022suppression}.
Finally, unlike prior studies on engineered dissipation~\cite{Stanningel_PRL2014, schmale2024stabilizing}, our approach leverages quantum trajectories, enabling a better reconstruction of the target dynamics and a deeper understanding of the role of measurements themselves.

By delving into the interplay between gauge theories and measurement-induced dynamics, we thus uncover insights into the out-of-equilibrium properties of highly constrained quantum systems, whose relevance extends far beyond LGTs and includes many classical optimization problems of practical relevance. 
Our study not only broadens the theoretical understanding of quantum dynamics in these intriguing systems but also expands the groundwork for the development of robust and efficient quantum simulation protocols and opens new pathways for the protection of quantum information encoded in noisy hardware.

The rest of the paper is organized as follows. In Sec.~\ref{sec:model}, we describe the $\Ztwo$ LGT we study to illustrate the error mitigation schemes.
We analyze the gauge protection in discrete-time dynamics in Sec.~\ref{sec:dps} and connect it to the quantum Zeno protection in analog quantum simulations. In particular, discuss the mapping emergence of a quantum Zeno transition in Sec.~\ref{sses:qzt} and analyze the different stochastic unravelings in Sec.~\ref{ssec:traj}.
We study the effect of imperfect measurements on the Zeno-protected phase in Sec.~\ref{sec:noise}.
In Sec.~\ref{ssec:ec}, we extend the DPS protocol to correct for the effect of incoherent bit-flip noise.
We discuss our results and future outlooks in Sec.~\ref{sec:conclusion}.
Further details and numerical results are presented in the Appendix.

\section{Model Hamiltonian}\label{sec:model}
To illustrate the interplay between measurements and gauge protection in the simplest way possible,  
we consider a 1+1d lattice gauge theory with $\Ztwo$ symmetry and dynamical matter \cite{Schweizer_NatPhys2019,halimeh2020fate,Mildenberger2025}. After a standard Jordan--Wigner transformation, the Hamiltonian $H_{\Ztwo} = H_J + H_f + H_m$ is the sum of three contributions
\begin{subequations}\label{eq:H}
   \begin{align}
H_J = & J\sum_n \left( \sigma^+_n \tau^x_{n,n+1} \sigma^-_{n+1} + {\rm h.c} \right) \ ,  \\
H_f = & f \sum_n \tau^z_{n,n+1} \ ,  \\
H_m =& \frac{\mu}{2} \sum_n (-1)^n \sigma^z_n \ .
\end{align} 
\end{subequations}

The Pauli operators $\sigma^\alpha_n$ act on the sites of a 1d lattice and represent the dynamical charges while the $\tau^\alpha_{n,n+1}$ are the $\Ztwo$ gauge field operators acting on the link between sites $n$ and $n+1$.
$H_J$ describes a nearest-neighbor particle hopping assisted by the gauge field connecting the two sites.
$H_f$ is an effective ``electric-field'' contribution to the Hamiltonian and $H_m$ is the staggered mass of fermionic degrees of freedom, which are interpreted as particles on even sites and antiparticles on odd ones.
We consider a system with open boundary conditions.

The Hamiltonian $H_{\Ztwo}$ commutes with the set of local symmetry operators ($\Ztwo$ gauge charges) $G_n = -\tau^z_{n-1,n}\sigma^z_n \tau^z_{n,n+1}$. 
Thus, the Hilbert space is split into subsectors labeled by the eigenvalues of $G_n$, i.e., an array of conserved local charges $\lbrace g_n\rbrace_{n=0,\dots, N-1}$, where $N$ is the number of lattice sites.

Since we are interested in possible error suppression and correction methods, we consider coherent as well as incoherent sources of errors. For the coherent errors, we chose spin flips of the gauge operators and unassisted tunneling between neighboring matter sites in order to represent typical local errors that may appear in a quantum simulation  \cite{Poppitz_2008,Mil_Science2020,Yang_Nature2020,Halimeh_PRXQ2021}. The corresponding Hamiltonian is 
\begin{equation}\label{eq:H_err}
    H_{\rm err}=\lambda_1\sum_n \tau^x_{n,n+1} + \lambda_2 \sum_n \left( \sigma^+_n \sigma^-_{n+1} +{\rm h.c.} \right) \ ,
\end{equation}
where $\lambda_{1(2)}$ is the strength of the corresponding error source. The Hamiltonian $H_{\rm err}$ does not commute with the symmetry generators $G_n$ and thus couples sectors with different values of the local charges. 
The normalized gauge violation
\begin{equation}\label{eq:gauge_v}
    \delta g = \frac{1}{2N}\sum_n |\langle G_n \rangle - g_{n,{\rm targ}}|
\end{equation}
monitors how far the error Hamiltonian is leading the state away from the target charge sector identified by the set of conserved charges $g_{n,{\rm targ}}$.
While other error sources can be considered, we do not expect their specific choice to qualitatively change the system dynamics, as long as $H_{\rm err}$ remains local and explicitly breaks the gauge symmetries.

We model incoherent errors as spin flips or phase flips (Pauli errors) randomly occurring during the time evolution. We will denote with $p_{\rm err}$ the probability per qubit per trotter step that one of these errors affects a quantum simulation.
As we are interested in harnessing the Gauss laws to detect and correct gauge-violating errors, we will consider only the Pauli-X channel that explicitly breaks the $\Ztwo$ gauge symmetries.

We use the python library qutip~\cite{qutip1,qutip2} for the numerical analysis of the following sections.
 
\section{Dynamical post-selection in digital quantum simulations.}\label{sec:dps}
\begin{figure*}
    \centering
    \includegraphics[width=2\columnwidth]{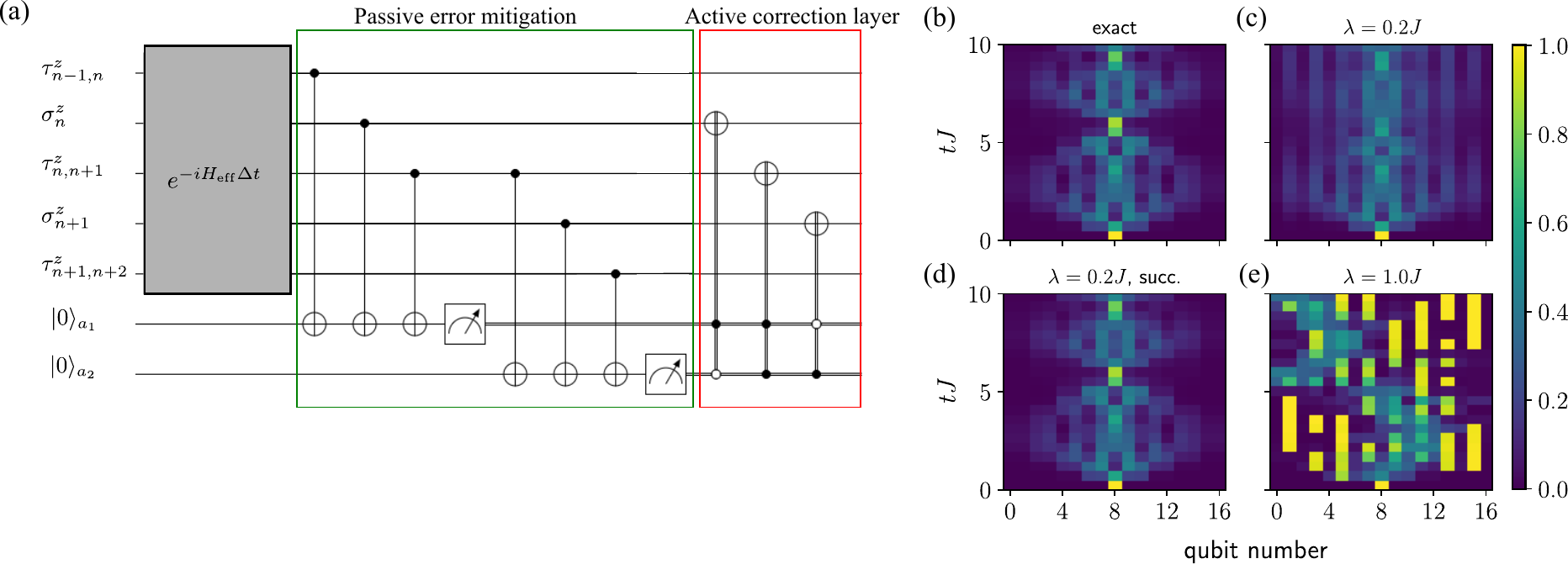}
    \caption{(a) Circuit representation of the DPS error mitigation/correction scheme. The gray box is a single Trotter step with an effective non-Hermitian Hamiltonian that includes gauge-breaking terms and, possibly, incoherent spin flips. The results of the measurements on the ancillary qubits have to be stored in a $N\times n_t$ classical register.
    The first block after the unitary evolution monitors possible gauge violations, mitigating coherent errors. The active correction layer, exerts feedback on the computational qubits based on the measurements on the ancillas, compensating for spin-flip incoherent errors. 
    (b-e) examples of dynamics affected by gauge-violating errors and measurements of the local gauge charges. The colormap represents the qubit excitation $(1-\sigma^z_n)/2$ for matter qubits and $(1-\tau^z_{n,n+1})/2$ for gauge variables.
    (b) Ideal trotterized dynamics. (c) Dynamics affected by coherent errors with $\lambda=0.2J$. The deviations from the ideal dynamics are clearly visible. (d) Single trajectory with successful dynamical post-selection in the Zeno regime. The ideal dynamics is restored to an excellent degree. (e) Irregular dynamics due to measurements out of the quantum Zeno regime where error suppression is no longer possible.
    All simulations have $L=17$ qubits ($N=9$ matter sites), electric coupling $f=0.5J$ and fermion mass $\mu=0$.}
    \label{fig:fig1}
\end{figure*}

We consider a digital quantum simulation protocol where the state evolves according to a standard first-order Trotter discretization of the time evolution operator
$\nep^{-iH t} = \left[ \nep^{-i H \Delta t}\right]^{n_t}$, with $t = \Delta t\, n_t$ being the total evolution time and $n_t$ the number of Trotter steps.
Each element is further decomposed in the product of simpler unitary operators generated by the single terms in the sums in Eqs.~\eqref{eq:H} and \eqref{eq:H_err}.
On top of this, we can consider incoherent noise effects by adding jump operators associated with dissipation or dephasing errors occurring with a certain probability within each time step.
We will later consider spin-flip errors in Sec.~\ref{ssec:ec}, where we employ a feedback correction mechanism. For now, we focus on the situation where the only error sources are coherent, with a single parameter $\lambda_1=\lambda_2=\lambda$ setting their strength.

To detect gauge violations arising during the time evolution, we couple the symmetry operators of the model, i.e., the local gauge charges, to ancillary qubits~\cite{BonetMonroig_PRA2018, Stryker_PRA2019,schmale2024stabilizing}, initially prepared in the fiducial state $|0\rangle_a$. 
Mid-circuit measurements on these auxiliary registers project the state either back in the target gauge sector, effectively removing the accumulated errors, or in an orthogonal sector.
When a gauge violation is measured, i.e., the ancilla state has flipped, the simulation is discarded as it gives an unphysical contribution to the evaluation of expectation values.
The outcome of each measurement is stochastic, despite the discretized evolution being deterministic, with the probability of their outcome depending on the amplitude of the gauge violation.
Importantly, we may dilute the measurements, either by measuring only a subset of local charges at each time step or by reducing the frequency of the measurements. 
This allows for the tuning of the effective protection strength.

The DPS circuit is exemplified in Fig.~\ref{fig:fig1}(a), where we show the error mitigation and the active correction layers acting on two neighboring local charges $G_n$ and $G_{n+1}$.
If all $G_n$ are measured at each Trotter step, $N$ ancillary qubits are needed. 
A reset of the ancillary qubits is not necessary as the simulation stays in the target sector as long as all ancillary qubits remain in their initial state $\ket{0}$. The outcome of each measurement, instead, can be easily stored in a number of classical registers growing linearly in both $N$ and the circuit depth.

To benchmark this method, we test it on a quantum quench corresponding to studying the spreading of an initially localized meson (consisting of a mobile charge and a fixed background charge) \cite{Mildenberger2025}. In this scenario, the system is initially prepared with the electric field all polarized in the $x$ direction, and a single mass excitation is set at the center of the chain. 
This state is then evolved with a standard first-order Trotter discretization of the evolution operator generated by $H=H_{\Ztwo}+H_{\rm err}$ with $\lambda=0.2J$, $f=0.5J$, and $\mu=0$. We consider a system with $L=17$ qubits, corresponding to $N=9$ local charges.
This choice corresponds to the deconfined phase of the $\Ztwo$ LGT where we expect the mass defect to spread ballistically in the chain and bounce off the edges, as the chain has open boundary conditions.
Indeed, this behavior is clearly visible in Fig.~\ref{fig:fig1}(b), where we show the local qubit excitation ($\frac{1-\sigma^z}{2}$ for site and $\frac{1-\tau^z}{2}$ for link variables) in such protocol with $\lambda=0$. 

When the coherent error is added, see Fig.~\ref{fig:fig1}(c), the correlation between matter and gauge variables is visibly smeared out. However, it is recovered completely with a successful application of the DPS protocol, shown in panel (d).
For the DPS to protect the gauge symmetries, the frequency of the measurements must be sufficiently high to prevent the error from increasing too much in a single Trotter step.
When the latter case happens, for instance, because the error strength $\lambda$ is too large, the protocol enters a non-protected regime where the measurements project the state onto a random gauge sector and the dynamics displays a very irregular behavior.
We report an example of this unprotected behavior in Fig.~\ref{fig:fig1}(e).

\subsection{Effective Lindblad master equation and quantum Zeno transition}\label{sses:qzt}

To understand the origin of these two regimes and determine the measurement frequency threshold for gauge protection, it is instructive to look at the continuous-time limit of the DPS protocol in the ensemble average over many trajectories.
An important aspect to consider is that the measurements in the DPS protocol are strongly correlated, as they are all performed at the end of a Trotter step.
In a continuous-time framework, we require instead independent measurements that do not occur simultaneously. 
A convenient workaround is to split the time step associated with the measurement frequency $\Delta t$ in smaller intervals $\Delta t = N_{\rm step} \ud t$ and assume that quantum jumps occur randomly in such finer steps with a probability $ \ud t/ \Delta t$.
Hence, $\Delta t$ becomes the average time in which $N$ charges are measured.

Let us momentarily neglect the unitary evolution; given a state $\ket{\psi(t)}$ of the system at a given time $t$, the measurement of a local charge $G_n$ leads to
\begin{equation}\label{eq:stochastic1}
    \ket{\psi(t+\ud  t)} = \left\{ \begin{array}{c}
         \frac{P_n^+\ket{\psi(t)}}{\sqrt{\bra{\psi(t)} P_n^+\ket{\psi(t)}}}  \ {\rm with}\ p_n^+= \frac{1+\langle G_n \rangle}{2} \ ,\\
         \frac{P_n^-\ket{\psi(t)}}{\sqrt{\bra{\psi(t)}P_n^-\ket{\psi(t)}}} \ {\rm with}\ p_n^-= \frac{1-\langle G_n\rangle}{2} \ ,
    \end{array} \right.
\end{equation}
where, according to the Born rule, $p_n^\pm=\bra{\psi(t)}P_n^\pm\ket{\psi(t)}$ is the probability of projecting on either one of the symmetry sectors and $P^\pm_n=\frac{1\pm G_n}{2}$ the corresponding projection operator.
The denominators  ensure the normalization of the state after a measurement and the expectation values are taken over the state $\ket{\psi(t)}$
\footnote{We note that the protocol \eqref{eq:stochastic1} defines a POVM since  $\left(P_n^+\right)^\dagger P_n^+ + \left(P_n^-\right)^\dagger P_n^- = 1$ and $p_n^+ + p_n^-=1$. In the expressions for the probabilities and normalizations we also exploited that $\left(P_n^\pm\right)^\dagger P_n^\pm = \left(P_n^\pm\right)^2=P_n^\pm$}.
Using Eq.~\eqref{eq:stochastic1} and taking into account the probability $\ud t/ \Delta t$ that a measurement occurs in the interval $\ud t$, we can write the evolution of the system density matrix $\rho(t)$ as 
\begin{eqnarray}
    \rho(t+\ud t) &=& \frac{ \ud t}{\Delta t} \left[ P^+_n\rho(t)P^+_n + P^-_n\rho(t)P^-_n \right] +\left(1-\frac{\ud t}{\Delta t}\right) \rho(t)  \cr
    &&\cr
    &=& \frac{\ud t}{\Delta t}\left[ 2 P^+_n\rho(t)P^+_n -\lbrace P^+_n,\rho(t) \rbrace \right] -\rho(t)\ ,
\end{eqnarray}
where we used $P^-_n=1-P^+_n$.
We can now expand at first order $\rho(t+\ud t) \simeq \rho(t) + \ud t \dot{\rho}(t)$ to derive an effective master equation of the density matrix, 
\begin{equation}\label{eq:stochastic2}
    \dot{\rho}(t) \simeq \frac{1}{2\Delta t} \left[ G_n \rho(t) G^\dagger_n -\frac{1}{2} \lbrace G^\dagger_n G_n, \rho(t) \rbrace\right] \ ,
\end{equation}
where we explicitly use the local charge $G_n$ instead of the associated projector.
A further simplification for the $\Ztwo$ gauge group follows from $G_n^\dagger=G_n$ and $G_n^\dagger G_n = G_n^2=1$, but the functional form of Eq.~\eqref{eq:stochastic2} makes it clearer that it corresponds to the contribution of a jump operator $G_n$ to a Lindblad master equation.
Inserting the unitary part and summing over all local charges, we finally get
\begin{equation}\label{eq:lindblad}
    \dot{\rho} = i[\rho,H] + \frac{1}{2\Delta t} \sum_n \left[ G_n \rho G^\dagger_n - \frac{1}{2}\lbrace G^\dagger_n G_n, \rho\rbrace \right] \ ,
\end{equation}
which is the Lindblad equation describing the evolution of the density matrix associated with the ensemble average of the stochastic trajectories originated by the measurement process.
We remark that the time step $\Delta t$ in Eq.~\eqref{eq:lindblad} is the average time between two consecutive measurements of the same local charge and not the Trotter step $\ud t$. 
When taking the continuous-time limit, $\ud t \to 0$, the effective measurement rate $\gamma \equiv \frac{1}{2\Delta t}$ is kept constant. That is, in this limit each generator is measured only once during many Trotter steps.  

From Eq.~\eqref{eq:lindblad}, we can draw a direct connection with previous proposals for gauge protection in analog quantum simulations~\cite{Stanningel_PRL2014}, which used random gauge rotations with vanishing auto-correlation time (white noise dephasing) to implement a master equation of the form of Eq.~\eqref{eq:lindblad}.
Both error mitigation mechanisms rely on the suppression of coherences between different gauge sectors due to the actions of the Lindblad operators $G_n$  in the large $\gamma$ limit.
In the context of measurement-induced dynamics, however, we can interpret Eq.~\eqref{eq:lindblad} as the ensemble average over stochastic quantum trajectories arising from the continuous weak measurements of the local charges $G_n$.
Within this protocol, until a quantum jump occurs, each trajectory evolves as 
$\ket{\psi(t)}=\exp(-i H_{\rm eff} t)/\mathcal{N}(t)$,  where we defined the effective non-Hermitian Hamiltonian
\begin{equation}\label{eq:Heff}
    H_{\rm eff} = H -\frac{i \gamma}{2} \sum_n G^\dagger_n G_n \ ,
\end{equation}
and $\mathcal{N}(t)$ ensures normalization of the quantum state.
Stochastic quantum jumps occur with uniform probability in time $\ud p = \gamma \ud t$ and modify the state according to  
\begin{equation}\label{eq:jump}
    \ket{\psi(t+ \ud t)} = G_n \ket{\psi(t)}/\sqrt{ \bra{\psi(t)} G^\dagger_n G_n \ket{\psi(t)}} \ .
\end{equation}
The total probability of a quantum jump event is thus $N\gamma dt$, where $N$ is the number of local charges (for the present theory equal to the number of matter sites).
The denominator ensures the normalization of the wavefunction; for $\Ztwo$ LGT, it is redundant, $G_n$ being unitary, but we write it explicitly to keep the notation more general.
In this context, $1/\gamma$ is the average time between consecutive jumps.
Notice, however, that it is twice as large as the average ```waiting'' time for the continuous limit of the DPS protocol, where $\Delta t = \frac{1}{2\gamma}$, a hint that the two approaches lead to different unraveling of the same Lindblad equation.
Indeed, also the measurement procedures are different: in the DPS protocol, we stroboscopically project the state on the target gauge sector while in the analog quantum simulation, we monitor the local charges with continuous weak measurements.
More details about the quantum trajectory approach are presented in App.~\ref{sec:qt}.

\paragraph{Symmetry-protection phase transition }

From the above and also previous discussions~\cite{Stanningel_PRL2014}, it may appear the error-mitigation regime emerges continuously for any value of the measurement rate with a protection strength proportional to $\gamma$.
Instead, we now show that the error-mitigation regime rather appears after a sharp quantum Zeno transition that signals the onset of the protected phase.
\begin{figure}
    \centering
    \includegraphics[width=\columnwidth]{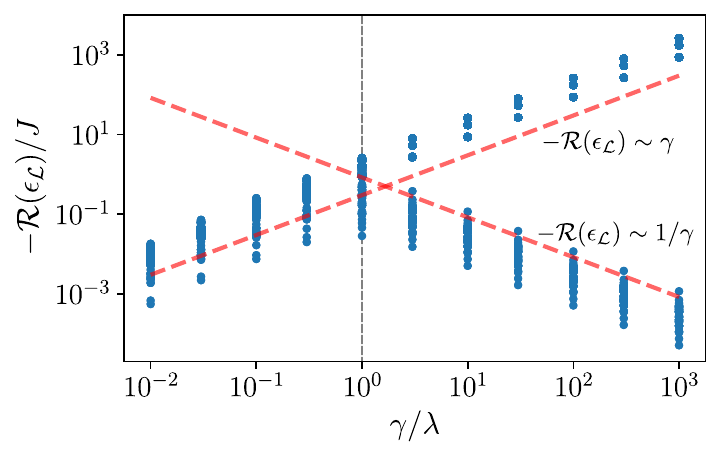}
    \caption{Absolute value of the real part of the Liouvillian spectrum for $N=3$ versus the measurement rate of the local charges, $\gamma$, and error strength, $\lambda=0.3J$.
    The remaining Hamiltonian parameters are $\mu=0$ and $f=0.5J$. The vertical dashed line is a guide to the eye marking the quantum Zeno transition. After it, two distinct groups of eigenvalues emerge, with opposite $\mathcal{R}(\epsilon_{\mathcal{L}})\sim \gamma^{\pm 1}$ highlighted by the dashed red lines.
    The lower group of eigenvalues (with $\epsilon_{\mathcal{L}}\sim 1/\gamma$) marks ``dark'' states that are stabilized by the measurement.}
    \label{fig:liouv_es}
\end{figure}

This transition is not manifested at the level of the stationary state of the associated Liouvillian operator $\mathcal{L}$, as it is always a trivial infinite-temperature mixed state.
Instead, it is revealed in the behavior of the real part of the nonstationary eigenvalues; these are all negative values whose amplitudes determine the decay rates for different eigenstates of $\mathcal{L}$. 
We plot the real part of the Liouvillian eigenvalues $\epsilon_\mathcal{L}$ in Fig.~\ref{fig:liouv_es}; the presence of a quantum Zeno point is signaled by their branching into two groups at $\gamma/\lambda =1$, marked by the vertical dashed line.
In the protected phase, $\gamma > \lambda$, a set of levels decay to zero as $1/\gamma$, consistently with the protection rate observed in Ref.~\cite{Stanningel_PRL2014}.
In the $\gamma \to \infty$ limit, the stochastic dynamics stabilizes these states that become ``dark" with respect to the jump operators. They correspond to all combinations \footnote{Eigenstates of $\mathcal{L}$ must be traceless but for the stationary state that is the only proper density matrix.} of eigenstates of $\lbrace G_n \rbrace $: 
Indeed, each gauge sector is equally protected by the Liouvillian dynamics, which is insensitive to the initial state. 
Hence, the dimension of the protected manifold is the same as the full Hilbert space. The jump operators destroy the coherence between states belonging to different sectors, thus preserving the initial configuration of local charges. 

\begin{figure*}
    \centering
    \includegraphics[width=2\columnwidth]{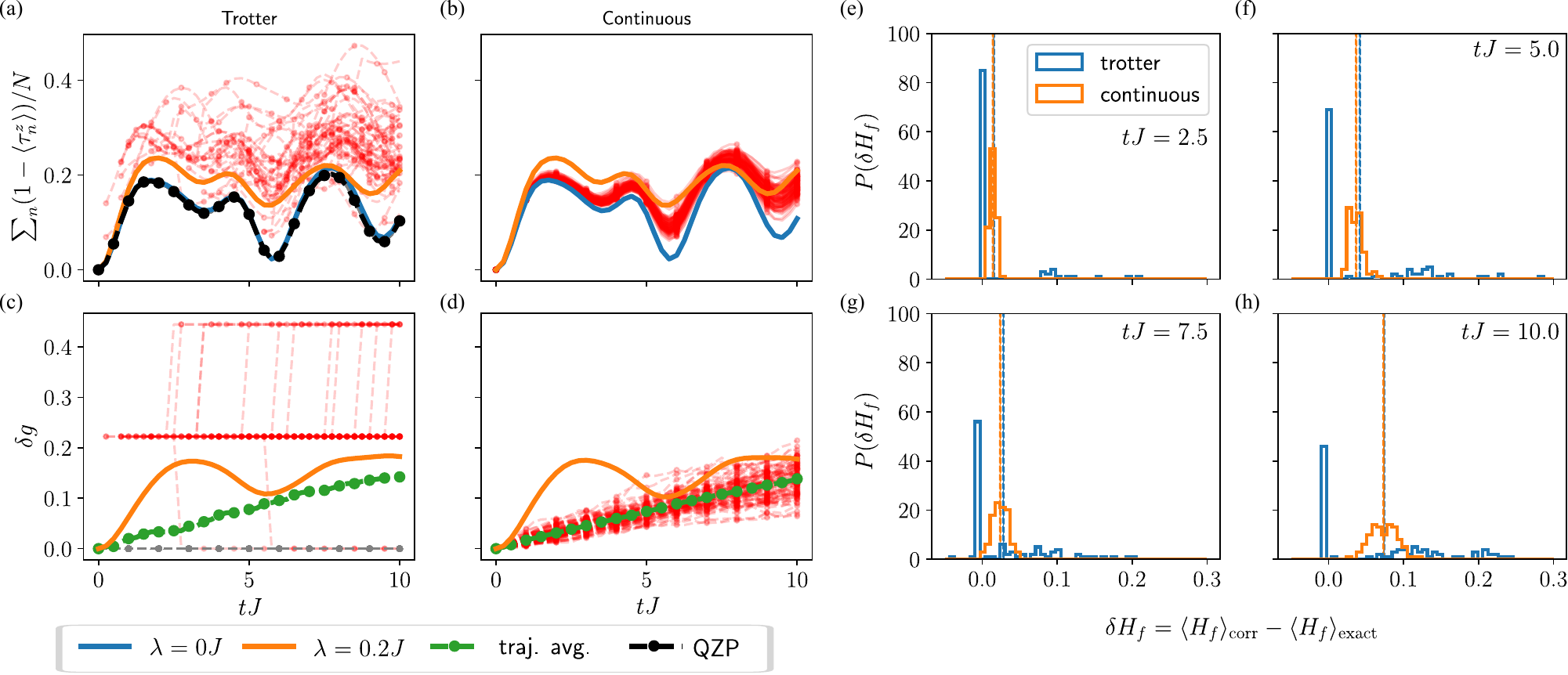}
    \caption{(a) Evolution of the average electric field in the deconfined phase, comparing the exact trotterized dynamics (blue), the coherent error (orange), and 100 trajectories with gauge measurements after each step (dashed lines). Each trajectory is depicted in gray as long as it does not break gauge invariance and in red from the first time a violation is measured. 
    The grey lines are all perfectly superimposed and their average (dashed black lines and circles) matches almost perfectly the exact evolution, as a consequence of quantum Zeno protection (QZP). The Trotter step is $\Delta t = 0.25 J^{-1}$
    (c) Evolution of the gauge violation, same protocols as in panel (a). 47 out of 100 trajectories never incur in a gauge violation.
    (b) and (d) plot the same observables as (a) and (c), respectively, obtained with a standard quantum trajectory unraveling of Eq.~\eqref{eq:lindblad}, with a rate $\gamma=\frac{1}{2\Delta t} = 2J$.
    By comparing panels (c) and (d), it is evident that the trajectory statistics are very different, despite their ensemble average is the same (green circles with dashed line).
    (e-h) Probability distribution of the error with respect to the exact dynamics in the expectation value of the electric field $H_f$. We compare the DPS protocol (blue) and the continuous weak measurements (orange).
    The normalized histograms are extracted from 500 independent trajectories for both protocols. The vertical dashed lines mark the average of the probability distributions.
    The averages of the two distributions are the same, despite their statistics differing significantly.
    The physical parameters of the Hamiltonian are $f=0.5J$, $\mu=0$, and $\lambda=0.2J$. The spin chain has $N=9$ matter sites (L=17 qubits).}
    \label{fig:trotter_suppression}
\end{figure*}

Armed with this knowledge, we can look again at the data presented in Fig.~\ref{fig:fig1}(d-e) to better understand the effect of the local measurements.
In both cases, the Trotter step is $\Delta t=0.5 J$, but it corresponds to different gauge protection strength depending on the error amplitude $\lambda$. In panel (d), $\gamma /\lambda =1/(2\Delta t \lambda) =5$, signaling that the protocol is in the protected regime of Fig.~\ref{fig:liouv_es}. Indeed, the coherent error is almost perfectly suppressed after each time step, allowing for the reconstruction of the exact dynamics.
In panel (e), in contrast, $\gamma/\lambda=1$, just below the transition to the Zeno phase. Consequently, the dynamics is highly irregular, with no resemblance with the ideal gauge-symmetric behavior.

In Appendix~\ref{app:Z3_and_U1}, we analyze the effective Liouvillian spectrum also for the $\mathbb{Z_3}$ and truncated $U(1)$ LGTs, showing that the phenomenology is qualitatively similar and independent from the specific symmetry group.

\subsection{Different unravelings for digital and analog gauge protection}\label{ssec:traj}

A natural question at this point is how the two stochastic processes compare with each other: On the one hand, we have the DPS protocol for discrete-time evolution, and on the other hand continuous measurements (or dephasing) of the local charges in an analog quantum-simulation setup. 
They are rooted in the same Liouvillian dynamics, meaning that the ensemble average is fully equivalent. As we show now, their unravelings in quantum trajectories show profound differences.

These are summarized in Fig.~\ref{fig:trotter_suppression}, where we investigate the same quench dynamics presented before, comparing trajectories for the two gauge protection mechanisms.
Figure~\ref{fig:trotter_suppression}(a-b) shows the time evolution of the electric field excitation $1-\tau^z_{n,n+1}$ averaged over all link degrees of freedom in 100 independent trajectories (faint dashed lines) obtained with the DPS protocol (panel a) and with continuous measurements (panel b).  
In both plots, we also report the exact evolution (solid blue line) of the average electric field as well as the dynamics in the presence of coherent errors with strength $\lambda=0.2J$ but no correction (solid orange line). 

In Fig.~\ref{fig:trotter_suppression}(a), the stochastic DPS trajectories are clearly split into two groups.
During the time evolution, a measurement of one of Gauss' laws may yield the wrong result $g_n \neq g_{n,{\rm targ}}$, incurring in a gauge violation, and the state is projected into the subspace orthogonal to the target sector. These trajectories are plotted as faint dashed red lines, starting from the time step at which the measurement fails. 
After the state is projected in the nonphysical sector, each trajectory behaves differently and their average tends to an infinite temperature state {\em orthogonal} to the target gauge subspace.
In several instances, however, the state is projected back into the target gauge sector throughout the entire time evolution (faint dashed grey lines). These trajectories perfectly collapse on each other and are, therefore, identical to their average (dashed black line with full cirlces), which follows the exact dynamics extremely closely.

The trajectory behavior in the continuous time evolution protocol is markedly different, as shown in Fig.~\ref{fig:trotter_suppression}(b). 
Individual trajectories now are much more concentrated around their ensemble average which drifts away from the ideal evolution on a timescale $\propto \gamma$ (see App.~\ref{sec:qt} for details), consistently also with the behavior of the Liouvillain eigenvalues in the protected phase.

The difference between the two stochastic processes emerges even more clearly when looking at the average gauge violation, defined in Eq.~\eqref{eq:gauge_v}, along individual trajectories, which we show in Fig.~\ref{fig:trotter_suppression}(c) and (d).
For the discrete-time evolution, gauge violation is quantized, as a failed local charge measurement leads to a jump in $\delta g$ corresponding to a flip of two neighboring generators $G_n$, $G_{n+1}$. 
When the gauge symmetry is protected, $\delta g=0$ does not increase (superimposed gray lines).
In contrast, in the continuous dynamics, each trajectory displays a similar slow increase of the average gauge violation, without any sharp jumps.
Despite their different distributions, however, once we average over the entire ensemble of stochastic dynamics, the mean gauge drift (green lines and circles) is the same.
Indeed, the trajectory averages in panels (c) and (d) are very similar~\footnote{In the ensemble average, we first compute the gauge violation $\delta g = \sum_n | \langle G_n \rangle -g_n|$ for an individual trajectory and then we average over many stochastic evolutions. Hence, the fact that the two protocols give the same average is not evident a priori.}.
Clearly, in DPS, the average only over trajectories where the error has been successfully suppressed remains totally flat, as no gauge violation occurs. An analog consideration in the continuous-time dynamics is not possible.

To further analyze the contrasting unravelings of the two protocols, we compare the error of the expectation value of the electric field $\langle H_f\rangle$ over 500 trajectories. The result is shown in Fig.~\ref{fig:trotter_suppression}(e-h) for four time snapshots.
As already observed, the trajectories for the trotterized dynamics (blue) clearly split into two groups with different behaviors, leading to a bimodal distribution. 
A single prominent peak corresponds to the simulations in which no gauge violation has occurred yet at time $t$, while a more irregular and broader part of the distribution describes all those states that have been projected on a space orthogonal to the target sector.
As time increases, weight is transferred from the sharp peak to the broad distribution, as more trajectories incur in at least one error during the measurement stage.
In continuous-time dynamics (orange), instead, the probability distribution remains centered around its average, while its width slowly increases.
The means of the two distributions are marked in each panel by a vertical dashed line of color matching that of the corresponding histogram. The two lines are almost superimposed, indicating that the ensemble average of the very different trajectory distributions is, indeed, the same.

\subsection{DPS survival probability}

The clear distinction in DPS between trajectories that have incurred gauge violation and those that have not permits us to postselect on states that reside within the physical subspace. This approach is indeed successfully pursued in current gauge-theory quantum-simulation experiments, albeit only for measurements on the final state \cite{Martinez_Nature2016,Mildenberger2025,Nguyen_PRXQ2022}. 
Despite the similarities, the two approaches are not equivalent.
In the standard post-selection, as the evolution time increases, we expect the number of trajectories satisfying the gauge symmetry to decrease exponentially in $t$, thus requiring a larger number of shots to obtain sufficient statistics within the physical subspace. 
Moreover, a final measurement of the local charges can not identify gauge violations that occurred at earlier times and then compensated by further errors later on.
In DPS, we still expect an exponential decrease of the number of successful runs but with two important differences with respect to standard post-selection.
First, within a successful run, the state can only drift out of the target sector between one measurement and the next one. If those are sufficiently close, we expect the probability of a second-order process that introduces an error through a virtual occupation of a nonphysical sector to be negligible.
Second, the decay rate of the successful trajectories will be dependent on the measurement frequency; in the strongly protected phase, where the quantum Zeno effect takes place, the population of the sector stabilized by the measurements decreases {\em linearly} in time.
For a better quantitative understanding. it thus becomes an important aspect to study the survival probability $P_s(t)=\left\langle\hat{P}_{\mathrm{phys}}(t)\right\rangle$, where $\hat{P}_{\mathrm{phys}}=\bigotimes_n P_n^+$ is the projector onto the physical subspace. Physically, $P_s$ measures the number of trajectories that never incur in a gauge violation as a function of time. 

As a rough estimate, the error probability for any measurement is of the order $p_{\rm err} \sim (\lambda \Delta t)^2$, as this is the rate at which other gauge sectors are populated in each Trotter step. Thus, the probability of surviving $n_t$ time steps for a system with $N$ local charges is approximately 
\begin{equation}\label{eq:p_succ}
P_s(t)  = (1-p_{\rm err})^{N n_t}\simeq \nep^{-(\lambda^2 \Delta t) N t},    
\end{equation}
where $t=n_t \Delta t$ is the evolution time.
It decreases faster for larger systems and longer evolution times, besides larger error strength, while it benefits from smaller Trotter steps. If $\Delta t$ is small enough, meaning that the frequency of measurements is faster than the growth rate of the error $\lambda^2$,
the dynamics reach a quantum Zeno regime where the exponential drift out of the target sector is well approximated by a linear dependence on time $P_s(t)  \simeq 1-(\lambda^2 \Delta t) N t$.
The decay rate is also connected to the Zeno timescale, corresponding to the simulation time up to which this linear approximation holds and we can extract reliably physical information from gauge-invariant trajectories. This Zeno timescale is approximately $t_{\rm Z} \sim (\lambda^2 \Delta t N)^{-1}$, meaning it increases linearly with the measurement frequency $\gamma=\frac{1}{2\Delta t}$.

\begin{figure}
    \centering
    \includegraphics[width=\columnwidth]{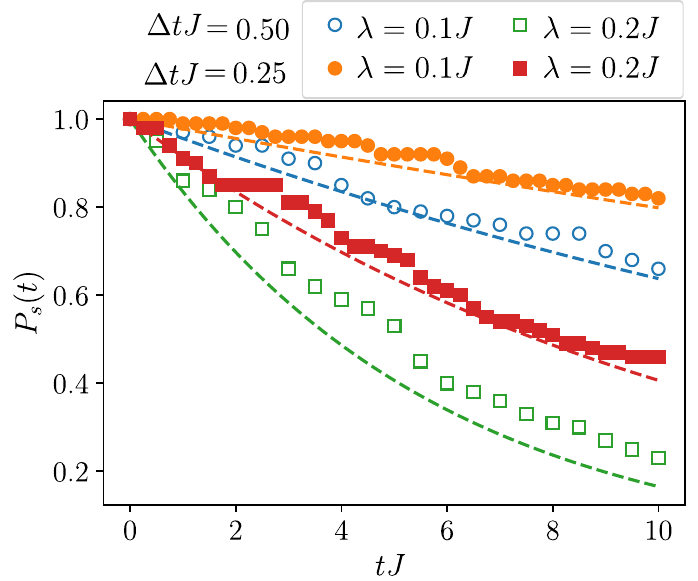}
    \caption{Number of trajectories that have not violated gauge invariance as a function of evolution time for different error strengths. Filled markers correspond to simulations with 40 Trotter steps ($\Delta t J=0.25$) while empty markers have $n_t=20$ ($\Delta t J =0.5$).
    The theoretical estimate, Eq.~\eqref{eq:p_succ}, for each dataset, is plotted with a dashed line with the same color as the numerical data. For small values of error strength and Trotter step size, the agreement is satisfactory and the data reproduces an approximate linear dependence at short simulated times.}
    \label{fig:p_survival}
\end{figure}
We plot this figure of merit in Fig.~\ref{fig:p_survival}, for two error strengths ($\lambda=0.1J$, $0.2J$; circles respectively squares) and two different Trotter steps ($n_t $=20, 40; empty respectively filled markers).
The success probability is extracted from 100 trajectories for each set of parameters.
We find a good agreement between the numerical data and  Eq.~\eqref{eq:p_succ} (dashed lines), in particular for small decay rates $\lambda^2 \Delta t$. 

In general, an exponential overhead is common to many error-mitigation techniques~\cite{Cai_RMP2023} and DPS makes no exception. Although this eventually limits their applicability on large systems, direct post-selection methods, for instance,  have already been successfully applied to study nontrivial dynamics in LGT experiments~\cite{Nguyen_PRXQ2022,Mildenberger2025,cochran2024visualizing}. 
Moreover, the accurate estimate of local observables only requires the system to satisfy the gauge symmetries within their light cone, allowing for extracting meaningful physical information also from trajectories that incur in a small number of gauge violations.
Since changing the monitoring frequency allows for tuning the decay rate in Eq.~\eqref{eq:p_succ}, our proposal can prove advantageous both in terms of sampling overhead and protection efficiency with respect to direct-post-selection, making it relevant for system sizes comparable with state-of-the-art NISQ devices ($\sim 100$ qubits), provided the mid-circuit measurements are sufficiently accurate.
In the next section, we investigate what happens if this is not the case.

\section{noisy measurements}\label{sec:noise}

Until now, we have assumed that the gates used to encode the local charge on the auxiliary qubits act with arbitrary precision, without introducing further errors or noise.
Despite highlighting the quantum Zeno transition more transparently, this is a significant simplification since two-qubit gates are the main noise sources in realistic hardware.

\begin{figure}
    \centering
    \includegraphics[width=\linewidth]{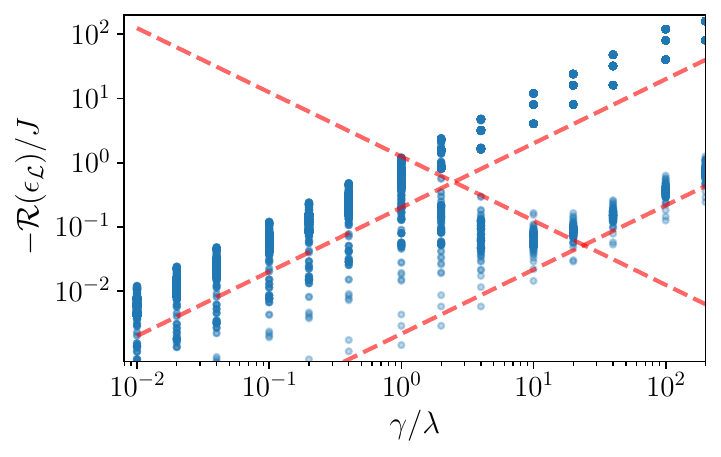}
    \caption{Magnitude of the real part of Liouvillian eigenvalues as a function of the ratio $\gamma/\lambda$ between the measurement frequency and the coherent error strength, for a measurement fidelity of $\mathcal{F}=0.99$. There remains an intermediate Zeno-protected phase. It is limited from below ($\gamma/\lambda\sim 1$) as before by the build-up of simulation errors when measurements are rare, and from above ($\gamma/\lambda\sim 10$) by large errors accumulating through the measurements themselves once they become too frequent. Marker opacity reflects the density of states and the dashed lines are guides to the eyes to highlight the three competing behaviors.
    }
    \label{fig:liouv-noisy}
\end{figure}
Here, we relax this assumption and model imperfect measurements through the action of random Pauli noise any time we couple a physical qubit with the ancilla.
Specifically, if we measure $G_n=-\tau^z_{n-1,n} \sigma^z_n \tau^z_{n,n+1}$, we also apply small random rotations around the $\bf{x}$ and $\bf{z}$ axes on the three involved system qubits, given by
\begin{eqnarray}
    U_{R,n}(\boldsymbol{\xi},\boldsymbol{\eta})= \bigotimes_{k \in \mathrm{supp} (G_n)} \nep^{-i\xi_k Z_k}\nep^{-i\eta_k X_k} \ .
\end{eqnarray}
\begin{figure*}
    \centering
    \includegraphics[width=\linewidth]{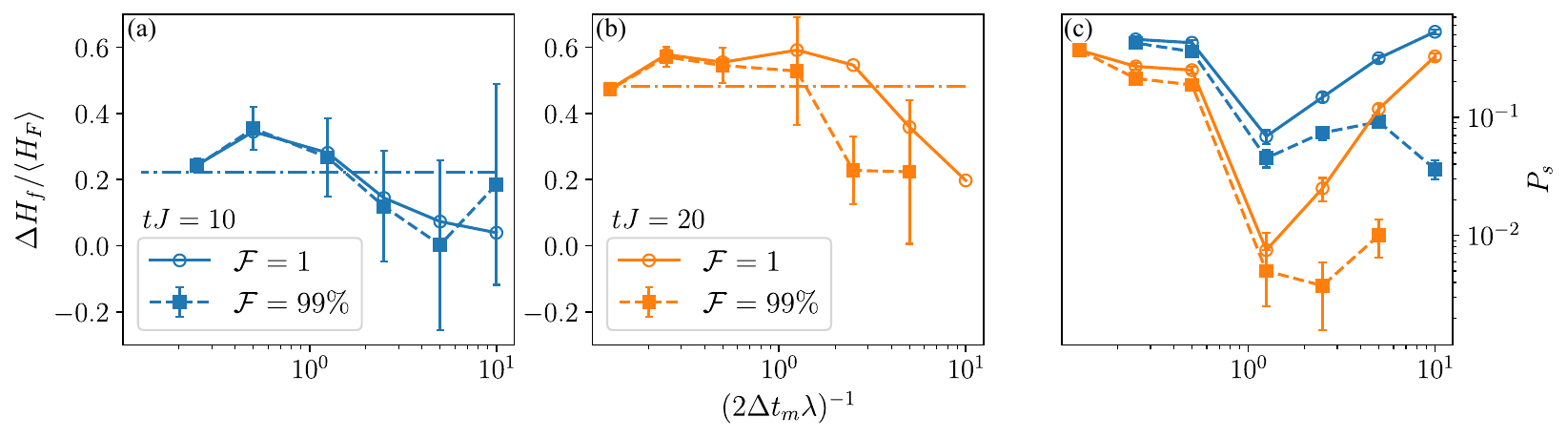}
    \caption{Relative error of the electric Hamiltonian at times $tJ=10$ (a) and $tJ=20$ (b) for DPS applied with varying measurement intervals $\Delta t_m$ and both ideal ($\mathcal{F}=1$) and imperfect ($\mathcal{F}=0.99$) measurements. We compare the data with a post-selection procedure only at the end of the simulation with no intermediate monitoring (horizontal dashed line). At sufficiently large measurement frequencies, DPS improves on top of direct post selection even with noisy measurement circuits.
    (c) Survival probability as a function of the measurement frequency. A dip appears in correspondence of the Zeno transition, with a worse recovery for finite infidelity $1-\mathcal{F}$. Data for $tJ=20$ and shortest (noisy) measurement period $\Delta t_m = \Delta t$ is missing because no physical DPS trajectory was found. Simulation parameters: $N=7$, $\Delta t J =0.25, f=0.5J$, $\lambda =0.2J$.}\label{fig:noisy_measurement}
\end{figure*}
$k$ labels the qubits in the support of $G_n$ and $Z=\sigma^z \ (\tau^z)$ when $k$ denotes a charge (gauge) degree of freedom, and analogously for $X$.
The angles $(\boldsymbol{\xi},\boldsymbol{\eta})$ of the rotations are drawn from a normal distribution with zero mean and a standard deviation $\sigma$ that sets the noise level.
$U_R$ maintains unitarity of the time evolution on a single trajectory but generates both dephasing and dissipation when we take the ensemble average.
We characterize its fidelity through the Hilbert--Schmidt inner product with the identity operator, averaged over noise realizations
\begin{eqnarray}
    \mathcal{F}=\overline{\left(\mathbb{1},U_R(\boldsymbol{\xi},\boldsymbol{\eta})\right)_\mathrm{HS}}\simeq 1-3\sigma^2 \ .
\end{eqnarray}
The effective error rate induced by noisy measurements is the infidelity $1-\mathcal{F}$.

A first nontrivial question regards the fate of the Zeno transition marked by the branching of the Liouvillian eigenvalues.
With finite measurement fidelity, the effective Lindblad equation approximately becomes
\begin{eqnarray}
        \dot{\rho} & = i[\rho,H] + \gamma \sum_n \left[ G_n \rho G^\dagger_n - \frac{1}{2}\lbrace G^\dagger_n G_n, \rho\rbrace \right] \nonumber \\
        & + (1-\mathcal{F})\gamma\sum_{k=1}^{2N-1} \left[ L_k \rho L^\dagger_k - \frac{1}{2}\lbrace L^\dagger_k L_k, \rho\rbrace \right] \ ,\label{eq:lindblad_noisy}
\end{eqnarray}
where $L_k = {X_k, Z_k}$ and $k$ runs over all $2N-1$ deegrees of freedom.
Hence, increasing the measurement frequency also increases the effective error rate.
If the eigenvalue branching remains, we expect the presence of an optimal region in the measurement frequency where the Zeno regime still exists. 
Figure~\ref{fig:liouv-noisy}, which reports the magnitude of the $\gamma$ dependence of the real part of the Liouvillian eigenvalues for $\lambda=0.2 J$ and $\mathcal{F}=0.99$,
confirms this picture.
The branching that signals the onset of the protected phase at $\gamma/\lambda = 1 $ is, indeed, still present but the measurement noise prevents the stabilization of the Zeno effect at large frequencies because of a bundle of eigenvalues growing as $-\mathcal{R}(\epsilon_\mathcal{L})\propto \gamma$ that ``bends'' upward the Zeno-protected states.
However, a finite range of measurement frequencies remains where the gauge protection from DPS is expected.

To confirm this prediction, we simulate the dynamics of a chain of $N=7$ charge sites ($13$ qubits) with the same quantum quench protocol as described in Sec.~\ref{sec:dps} and varying measurement frequency. In practice, we fix the total evolution time $t_f J =20$ and the Trotter step $\Delta t J =0.25$ but we dilute the measurements. We denote with $\Delta t_m$ the time interval between consecutive measurement layers, such that the monitoring frequency becomes $\gamma=\frac{1}{2 \Delta t_m}$.
We take  $\Delta t_m$ as an integer multiple of the Trotter step $\Delta t$.

Figures \ref{fig:noisy_measurement} (a) and (b) report the relative error of the electric field Hamiltonian at two time slices $tJ=10$ (a)  and $tJ=20$ (b), averaged over successful DPS trajectories out of 800 independent runs.
We compare ideal measurements ($\mathcal{F}=1$, circles) with noisy ones ($\mathcal{F}=0.99$, squares) and a direct post-selection performed only at the end of the run (horizontal dashed line). The latter corresponds to a measurement period of $\Delta t_m = tJ$.
When the measurements are perfect, increasing their frequency first leads to worse performances with respect to direct post-selection. After the Zeno transition ($2\Delta t_m\lambda \simeq 1$), the error starts decreasing and, eventually, DPS performs better than direct post-selection.
Interestingly, noisy measurements do not significantly alter this picture and occasionally even lead to better results than ideal DPS. The fluctuations, however, markedly increase for larger measurement frequencies as the accumulated errors tend to diffuse the trajectories around their average while, at the same time, reducing the number of successful ones.
At small $\Delta t_m$, noise overshadows the Zeno effect and the error starts growing again.

The survival probability also reflects the presence of the Zeno transition and its persistence at a finite error rate in the measurements, as is visible from Fig.~\ref{fig:noisy_measurement}(c).
It shows the survival fraction corresponding to the data in panels (a) and (b) as a function of the measurement frequency. At the Zeno critical point, $P_s$ has a clear dip followed by a sharp increase that surpasses the ratio of trajectories retained by direct post-selection (leftmost points).
The noisy measurements affect $P_s$ in the Zeno regime, reducing the number of trajectories that can be used for reconstructing the target dynamics.
In particular, the survival probability in Eq.~\eqref{eq:p_succ} is modified with an effective error rate that grows with $(1-\mathcal{F})$
\begin{eqnarray}
    P_s(t)  \simeq \nep^{-[\lambda^2 \Delta t_m+(1-\mathcal{F})/\Delta t_m] N t}. 
\end{eqnarray}
Importantly, now it is no longer possible to arbitrarily reduce the exponential decay rate of $P_s$ by making more frequent measurements but one has to balance the benefit of the Zeno-like stabilization with the additional noise.

To summarize, imperfect measurements have two main effects: first, they induce an effective relaxation within the target gauge sector, reducing the quality of the reconstructed dynamics. Second, the fraction of successful DPS trajectories decreases due to the higher error rate. 
Thus, while it is no longer possible to arbitrarily improve the gauge protection by increasing the measurement density, an optimal frequency range exists close to the quantum Zeno transition where mid-circuit measurements still provide the desired mitigation against coherent errors. In this regime, our protocol leads to a better approximation of the target dynamics than standard post-selection, as it suppresses multiple gauge violations that combine in a gauge-invariant error.
The existence of the optimal region in the measurement frequency depicted in~\ref{fig:noisy_measurement} does not depend on the gauge group. In Appendix~\ref{app:Z3_and_U1}, we report the same phenomenon for the gauge groups $\mathbb{Z}_3$ and $U(1)$.

\section{Error correction}\label{ssec:ec}
As we have seen in Sec.~\ref{sec:dps}, reading out the violations of gauge symmetry at each time step suppresses errors due to a quantum Zeno effect. In addition, it also opens up the possibility to actively correct the measured errors. 
This active correction scheme is represented by the full circuit sketched in Fig.~\ref{fig:fig1}(a). 
Its implementation requires on-chip feedback during the quantum algorithm. Such a feature is currently actively researched due to its importance for general error correction and is expected to become available in the next generation of quantum chips~\cite{Kerckhoff_PRL2010,Cramer_NatComm2016,Battistel_2023}. We will see below that one can exploit it for significantly improving the quality of gauge theory quantum-simulation data, without the requirement for full error correction capabilities. 

First, we test the correction capabilities on a bit-flip (Pauli-X) channel, where we only consider the incoherent errors that the local charge measurements can correct without recurring to more sophisticated QEC protocols~\cite{Rajput_npjQI2023,spagnoli2024faulttolerant}.
For simplicity, here we resort again to ideal measurements. Alternatively, one could consider a larger error rate incorporating the effect of the two-qubit gates used in the coupling with the ancillas.
We model the bit-flip error by assuming that, during each Trotter step, an incoherent spin flip can occur on any qubit with probability $p_{\rm err}$.
These errors can be detected by local charge measurements by comparing them with the known target sector~\cite{Stryker_PRA2019, BonetMonroig_PRA2018}, using the same circuit as sketched in Fig.~\ref{fig:fig1}(a).
Then, we apply the reverse bit-flip operation and correct the error depending on which gauge symmetries have been violated.
This procedure works unambiguously only when the error probability is sufficiently small such that the probability of two errors occurring in the same Trotter step is negligible. 
Indeed, when two bit-flips affect neighboring local charges, a simple measurement can not determine which qubits were affected by the dissipation errors.
In our numerics, when errors with ambiguous detection happen, we discard the run and do not use it to compute the expectation values of observables of interest.
\begin{figure}[t]
    \centering
    \includegraphics[width=\columnwidth]{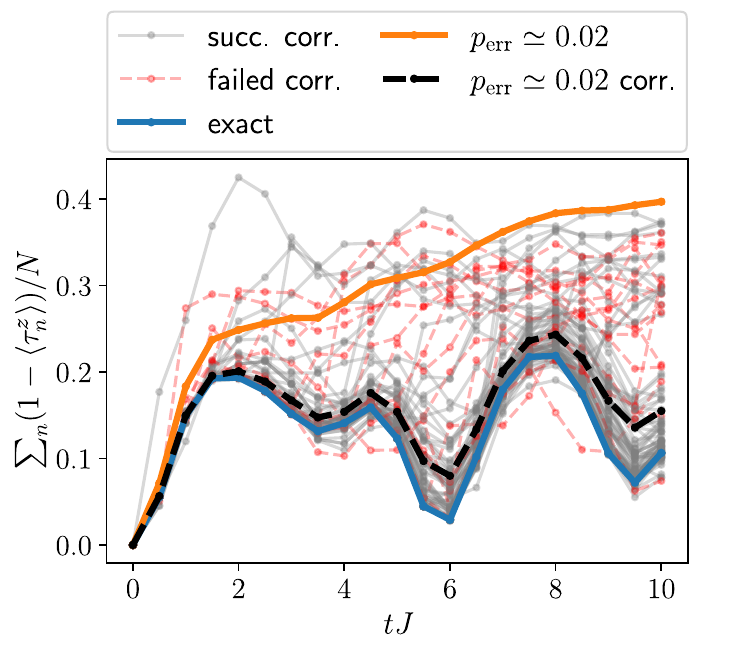}
    \caption{Example of error correction effectiveness with a pure bit-flip error channel. The orange solid line reports the uncorrected expectation value averaged over 100 noisy runs. The dashed black line is the average over the successfully corrected trajectories (faint grey lines), while we discard the trajectories where the error correction can not be applied unambiguously (faint dashed red lines). The corrected result gets very close to the exact data (blue solid line). }
    \label{fig:correction_only_flip}
\end{figure}

We report an example of the correction scheme in Fig.~\ref{fig:correction_only_flip}, where we compare the exact dynamics (solid blue line), the average over 100 noisy trajectories (solid orange line), and the average over successfully corrected trajectories (dashed black lines).
The gauge symmetry protection is not as effective as the dynamical post-selection in the presence of coherent error sources (see Fig.~\ref{fig:trotter_suppression}(a) for comparison). However, it is able to recover qualitatively the exact dynamics.
This feature is noteworthy,  considering the different type of noise (here: incoherent dissipation) and the completely deviating behavior of the noisy trajectories average.

When looking at individual quantum trajectories, however, it is apparently no longer possible to qualitatively distinguish those that have been unambiguously corrected (grey) and those that have not (dashed red).
Indeed, while there is a bundle of corrected trajectories that closely follows the exact dynamics, there are some in which the correction is apparently applied with success but they result in an irregular behavior similar to those where the correction can not be performed.
These originate when two neighboring gauge symmetries are violated by two independent spin flips on the matter sites instead of a flip of the gauge field, which is more probable and therefore implicitly assumed in the correction scheme.
This situation occurs with a smaller probability than single bit-flips, hence the average is dominated by the group of trajectories where the correction is successful.
As time increases, more trajectories will incur in such non-detectable double spin-flips, reducing the effectiveness of this error correction scheme for long evolution times.

Hence, it is clear that estimating the probability of nondetectable errors is of great importance in verifying the effectiveness of the correction scheme.
For instance, let us consider $G_n$ and $G_{n+1}$, which involve 5 qubits; if errors affect two of them, the correction protocol fails.
Given the error probability $p_{\rm err} \ll 1$ on a single qubit, the probability of having two errors in the same Trotter step affecting spins belonging either to $G_n$ or $G_{n+1}$ approximately reads
\begin{equation}
    p_2 = p_{\rm err}^2(1-p_{\rm err})^{L-2}6(N-1) \ ,
\end{equation}
where $L$ is the total number of qubits (matter and gauge sites), while $N$ is the number of matter sites. The factor $6(N-1)$ is the number of qubit pairs where errors can occur and affect two consecutive local charges. In this estimate, we neglect the effect of the boundaries.
Neglecting the possibility that three or more consecutive errors happen in the same Trotter step, the probability that a simulation fails due to these non-correctable errors is, therefore, $1-P_{\rm corr}=1-(1-p_2)^{n_t}$.
For the data presented in Fig.~\ref{fig:correction_only_flip}, where $p_{\rm err} = 0.02$, $L=17$, and $n_t=20$, this estimate gives a total error probability at the end of the time evolution of $P_{\rm err}=1-P_{\rm corr}\simeq 0.25$.
This estimate is compatible with the numerical data: out of 100 trajectories, 16 are discarded because of ambiguous errors (dashed red lines), and 15 lead to wrong results because of nondetectable double errors (outlier grey lines).

To appreciate the effectiveness of the correction scheme, we compare it to the survival rate of a noisy evolution, where after $n_t$ Trotter steps the probability that no incoherent bit-flip affected the dynamics is $P_{\rm succ} =(1-p_{\rm err})^{Ln_t}$. 
The ratio between the two success probabilities is
\begin{eqnarray}
    \frac{P_{\rm corr}}{P_{\rm succ}}= \left[ \frac{1-p_{\rm err}^2(1-p_{\rm err})^{L-2}6(N-1)}{(1-p_{\rm err})^L}\right]^{n_t} \ .
\end{eqnarray}
Assuming that $p_{\rm err}<1/L$ is small, a first-order expansion of the error rate per Trotter step leads to 
\begin{eqnarray}
    \frac{P_{\rm corr}}{P_{\rm succ}} \sim (1+Lp_{\rm err})^{n_t} \ .
\end{eqnarray}
The error correction scheme gives an {\em exponential} protection as $n_t$ increases, with respect to the bare noisy evolution, with a larger ratio for larger error probabilities.

Finally, let us combine the two types of error sources we discussed so far, i.e., local incoherent spin flips and the coherent error driven by the Hamiltonian in Eq.~\eqref{eq:H_err}.
Now, enforcing gauge protection becomes more complicated as ancilla measurements detect both incoherent spin flips and coherent gauge drift but cannot discriminate between them. A correction with a further bit flip is attempted in both cases but successfully compensates only the former. Correlated coherent errors cannot be inverted, indeed, by a single local bit-flip. 
The suppression of the coherent gauge drift, therefore, relies on the ``passive'' protection offered by the quantum Zeno effect through frequent measurements alone.

We adopt the same protocol as discussed so far, with both DPS and bit-flip correction, on top of which we implement a simple post-selection of the corrected trajectories.
To eliminate automatically those instances where the correction failed, we exclude the erratic trajectories far from the exact dynamics.
First, we evaluate the mean and standard deviation (std) of an observable, the electric Hamiltonian $H_f$ for instance, using all the simulations in which the gauge violation has been corrected.
Then, we exclude all the data that wander sufficiently far from the average, assuming they are affected by undetectable errors. 
From the remaining trajectories, we extract a new ensemble average and variance.
We iterate this procedure twice, first by excluding data that differ more than one standard deviation from the average, then those that differ more than two {\em updated} std from the new mean.
We present more details on this technique in App.~\ref{app:filter}.
It is important to remark that this post-processing may fail in actual experiments as it requires an accurate estimate of the expectation value of an observable on a single trajectory. 
Nevertheless, its success suggests that more refined clustering methods might produce similar effects on data collected with projective measurements on the computational basis.
\begin{figure}
    \centering
    \includegraphics[width=\columnwidth]{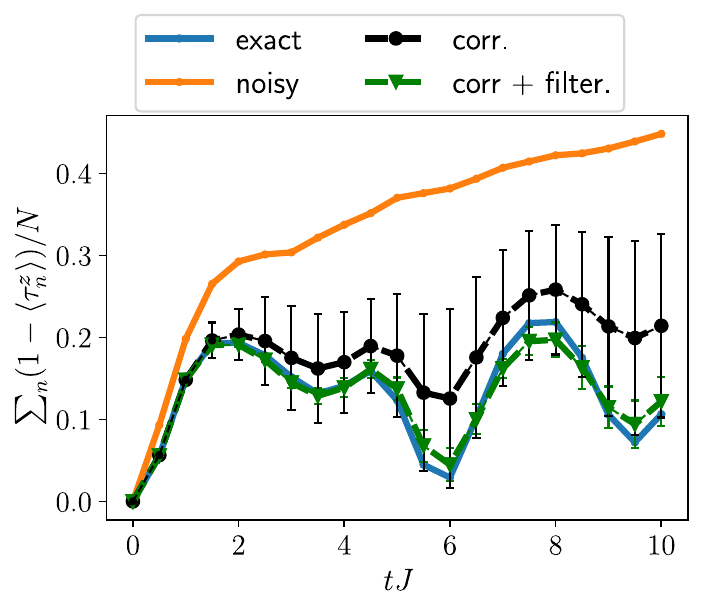}
    \caption{Active error correction. The noisy dynamics (solid orange line) deviates significantly from the exact dynamics (solid blue line). The average over corrected trajectories without the additional filter (dashed black line and circles) reaches considerably closer to the exact dynamics.
    An additional filter that iteratively discards outlier trajectories reproduces the exact solution to an excellent degree (dashed green line and triangles). The incoherent error probability per Trotter step at each qubit is $p_{\rm err}=0.02$ and the coherent error strength is $\lambda=0.2J$.}
    \label{fig:dynamics_full_error}
\end{figure}

Figure \ref{fig:dynamics_full_error} exemplifies the effectiveness of this method; while the DPS with EC already reduces the overall error from the noisy dynamics (dashed black and solid orange lines respectively), the additional filtering of trajectories leads to excellent protection of the dynamics from both coherent and bit-flip errors (dashed green line).
Moreover, the filtering also reduces the trajectory dispersion, represented by the error bars, selecting only those following the exact evolution (solid blue line).
While the trajectory variance with DPS is comparable with the electric field shot noise, after the filter their dispersion is greatly reduced, allowing for a more precise estimate of the expectation value.

A key take-home message is that within the framework of mid-circuit measurements, which naturally occur in any error-correction scheme, both coherent and incoherent errors can be handled. 
The efficiency of such protocols is limited by the measurement accuracy and the consequent maximum monitoring frequency. Understanding the intricate interplay between these many ingredients is a crucial step to bridge current NISQ devices to the primitive fault-tolerant ones, where the measurement-induced effects and imperfect error detection could still affect the circuit dynamics in nontrivial ways.

\section{Conclusions}\label{sec:conclusion}
In this paper, we have proposed a dynamical post-selection protocol harnessing passive mid-circuit measurements to both suppress (coherent) and correct (incoherent) violations of local gauge symmetries in quantum simulations of Abelian LGT models.
The error mitigation capacity is not continuous in the monitoring frequency but emerges only after a quantum Zeno transition in the spectrum of an associated Liouvillian superoperator, which separates the protected phase from a regime in which the measurements induce a dynamics that quickly thermalizes to an infinite temperature ensemble.
This framework based on the quantum Zeno effect unifies many different proposals that exploit stochastic dynamics to protect the gauge symmetries in both digital and analog quantum simulation protocols, such as random gauge transformations~\cite{Lamm_arxiv2020}, engineered dissipation~\cite{Stanningel_PRL2014}, and syndrome measurements~\cite{Stryker_PRA2019}. 
Crucially, despite the underlying Liouvillian being identical, each protocol results in a different unraveling in quantum trajectories. 
We exemplified this feature by comparing the stroboscopic projective measurements in a digital quantum simulation (the DPS protocol), with continuous weak measurements of the local charges in a continuous-time setup.
The ensemble averages of the two stochastic processes are the same while their distributions are surprisingly different.
Importantly, noisy measurements do not spoil the gauge protection, although the Zeno-like regime does not hold any longer to arbitrary large monitoring frequency.

Moreover, we showed that a simple error-correcting scheme harnessing the natural redundancy in LGTs~\cite{Rajput_npjQI2023} can be easily integrated with DPS circuits. 
It relies purely upon the constraint imposed by the local gauge symmetries to detect and correct the incoherent bit-flip errors with a feedback mechanism.
Though weaker than similar fully error-correcting schemes~\cite{Rajput_npjQI2023,spagnoli2024faulttolerant}, it is also significantly less demanding as it only requires at most $N$ ancilla qubits for the measurements, instead of auxiliary computational qubits and the associated heavier logical gates. 
It showcases how coherent and incoherent errors are handled differently within the same quantum circuit: first, frequent measurements suppress the coherent gauge drift, and, second, the active feedback builds upon the measurement results to compensate for incoherent spin-flips.
Hence, there is no need to implement further gauge protection schemes to suppress coherent errors, typically also requiring entangling gates, if mid-circuit measurements are sufficiently reliable.
Considering that the exponential overhead of error mitigation techniques limits their scalability in NISQ devices, it is of fundamental importance to understand how far we can push the present generation of quantum technologies in the quest for fault tolerance.
Our approach provides valuable insights into the transition between NISQ and fault-tolerant devices, where quantum resources are still limiting factors and error-mitigation/correction strategies also need to cope with their possibly undesired side effects on quantum simulations.
It stands as an important open research question whether error-mitigated NISQ devices can provide a real advantage for practical problems or give a precious contribution to guide the development towards more advanced scalable algorithms.

In our analysis, we focused on the $\Ztwo$ gauge group in 1+1 dimensions to illustrate our scheme in the simplest scenario. The extension to larger spacetime dimensions is conceptually trivial, even though the measurements of the local charges involve a larger number of entangling gates to map their parity to the ancillary qubits~\cite{Stryker_PRA2019,Ballini_2024}. 
To consider different abelian LGTs with larger groups, we studied the effective Liouvillian spectrum associated with the Zeno transition in models in one spatial dimension with $\mathbb{Z}_3$ and $U(1)$ symmetries, the latter truncated to a quantum link model with $S = 1$~\cite{Chandrasekharan_NucPhysB1997,Wiese_AnnPhys2013,Martinez_Nature2016,Surace_PRX2020,Nguyen_PRXQ2022}. 
As shown in Appendix~\ref{app:Z3_and_U1}, we obtained qualitatively similar results with $\Ztwo$ LGT, even considering noisy measurements.

When extending our protocol to larger groups, the main challenge is to design efficient measurement and correction circuits. 
Errors are still detected by monitoring the local charges~\cite{Stryker_PRA2019} but inverting the errors becomes more complicated because of the richer structure of possible gauge-breaking terms.
In this situation, a natural platform choice is represented by qudits, where a single quantum register encodes either the full symmetry group of a link or an appropriate truncation~\cite{GonzalezCuadra_PRL2022,Popov_PRR2024,Calajo_PRXQ2024}.
As the local charges adopt more values, the ancillary registers must also be encoded in qudits with a sufficient number of levels~\cite{Ballini_2024}. 
These features complicate the design of the DPS circuit but do not change its core idea, although, for non-Abelian groups,  results may depend on the order in which the local symmetry operators are measured.
On a positive note, using qudits would reduce the number of entangling gates needed for the symmetry verification, with respect to qubit devices, thus making the impact of measurement-induced noise less severe. 
Larger groups also impose stricter constraints on the physical states, meaning that monitoring the local charges would suppress a broader range of errors.

An aspect that we did not consider in this paper is the effect of a dephasing error channel. Indeed, we showed that DPS with error correction can suppress both coherent and bit-flip gauge-breaking errors, but it is insensitive to errors not affecting Gauss' laws.
A possible approach to refine our proposal is to use qubit redundancy with majority votes to correct dephasing errors and the local charge measurements for bit-flips~\cite{Rajput_npjQI2023,spagnoli2024faulttolerant}, at the price of implementing more costly error-corrected logical gates.
Moreover, in our analysis, we assumed that the error accumulated in each layer of the circuit is proportional to the discretization step $\Delta t$. 
Although this is irrelevant in the DPS scheme, it will likely affect its performance dependence on the number of Trotter steps.
When considering the error rates of nonparametric gates, such as CNOT, the Zeno protection will appear only in an optimal window of measurement frequency ranges. This is essentially what emerges when the circuit coupling the ancilla with the computational qubits is noisy, as we have shown in Sec.~\ref{sec:noise}, resulting in an imperfect but still effective error suppression.
In the future, it will be interesting to perform a thorough analysis with realistic error and noise sources to fully assess the efficiency of DPS in practical scenarios.

To conclude, thanks to the steady improvements in quantum computing and simulation platforms, mid-circuit and partial measurements are likely to become fundamental tools for manipulating and monitoring quantum systems~\cite{Kerckhoff_PRL2010,Cramer_NatComm2016,Battistel_2023,Pino_Nature2021,hashim2024quasiprobabilistic}.
Reaching a deeper understanding and control of measurement-induced effects such as quantum Zeno dynamics is, therefore, of great importance for the future developments of both quantum hardware and software.

\section*{Acknowledgements}
We thank J.~Mildenberger, L.~Spagnoli, A.~Russomanno,  and M.~Burrello for fruitful discussions.
This project has received funding from the European Union’s Horizon Europe research and innovation programme under grant agreement No 101080086 NeQST and from European Union - NextGeneration EU, within PRIN 2022, PNRR M4C2, Project TANQU 2022FLSPAJ [CUP B53D23005130006].
P.H. has further received funding from the Italian Ministry of University and Research (MUR) through the FARE grant for the project DAVNE (Grant
R20PEX7Y3A), from the Swiss State Secretariat for Education, Research and Innovation (SERI) under contract number UeMO19-5.1 , from the QuantERA II Programme through
the European Union’s Horizon 2020 research and innovation programme under Grant Agreement No 101017733, from the European Union under NextGenerationEU, PRIN 2022 Prot. n. 2022ATM8FY (CUP: E53D23002240006), from the European Union under NextGenerationEU via the ICSC – Centro Nazionale di Ricerca in HPC, Big Data and Quantum Computing.
This project has been supported by the Provincia Autonoma di Trento and Q@TN, the joint lab between the University of Trento, FBK—Fondazione Bruno Kessler, INFN—National Institute for Nuclear Physics, and CNR—National Research Council.
Views and opinions expressed are however those of the author(s) only and do not necessarily reflect those of the European Union or the European Commission. Neither the European Union nor the granting authority can be held responsible for them.
AB would like to thank the Institut Henri Poincaré (UAR 839 CNRS-Sorbonne Université) and the LabEx CARMIN (ANR-10-LABX-59-01) for their support.

\section*{Data availability statement}
All data supporting this work can be found in \cite{wauters_2024}.

\appendix
\section{Continuous measurement of local charges.}\label{sec:qt}
In the following, we give more details on how continuous measurements of the local charges can suppress the gauge violation induced by $H_{\rm err}$ in analog quantum simulations.
Our starting point is the Lindblad equation studied in Ref.~\cite{Stanningel_PRL2014}. In that work, it was shown that adding dephasing terms proportional to the Gauss' law generators leads to a suppression of the gauge-breaking error that scales as $1/\gamma$, where $\gamma$ is the effective rate determining the strength of the incoherent part of the dynamics. The full Lindblad equation is given by   
\begin{equation}\label{eq:app_Lindblad}
    \dot{\rho} = i[ \rho, H_0 + H_{\rm err} ] + \gamma \sum_n G_n \rho G^\dagger_n - \frac{1}{2} \lbrace G^\dagger_n G_n, \rho \rbrace \ .
\end{equation}
Equation \eqref{eq:app_Lindblad} describes a ``mean'' evolution of the density matrix, i.e., it represents an ensemble average over (infinitely) many realizations of the time evolution of the system coupled to some environment through the Lindblad (jump) operators $G_n$ with a common effective rate $\gamma$. 
Hence, a direct solution of Eq.~\eqref{eq:app_Lindblad} only describes the evolution of the expectation values of the observables over a statistical ensemble while it tells us nothing about what happens at individual trajectories of the quantum system or, in other words, at a single experimental realization of such a protocol.
In particular, a question we address is whether the gauge violation is suppressed by the action of the jump operators in each simulation or if it emerges only after averaging over many trajectories.

To study the evolution of states, instead of density matrices, under open dynamics, a possibility is to {\em unravel} the Lindblad equation in quantum trajectories. The first step is to incorporate the second term of the incoherent contribution into an effective non-Hermitian Hamiltonian
\begin{equation}\label{eq:app_Heff}
    H_{\rm eff} = H -\frac{i \gamma}{2} \sum_n G^\dagger_n G_n \ .
\end{equation}
The non-Hermitian part of the effective Hamiltonian causes an effective reduction of the norm of the wavefunction $\ket{\psi(t)}$; at first order in a time step $dt$, we have that 
\begin{align}
  \bra{\psi(t+ \delta t)}\psi(t+\delta t )\rangle &\simeq 1-\delta t \gamma \sum_n \bra{\psi(t)} G^\dagger_n G_n \ket{\psi(t)}\nonumber \\ 
  &= 1-\delta p \ ,
\end{align}
where $\delta t$ must be chosen such that $\delta p \ll 1$. The latter is then interpreted as the probability that the state undergoes a quantum jump described by 
\begin{equation}\label{eq:jump_app}
    \ket{\psi(t+ \delta t)} = G_n \ket{\psi(t)}/\sqrt{\gamma \bra{\psi(t)} G^\dagger_n G_n \ket{\psi(t)}} \ .
\end{equation}
The denominator is used to keep the normalization of the wavefunction.
This approach gives rise to stochastic dynamics in which the jump operators, in this case the generators of the local $\Ztwo$ symmetry, act at a random time with a probability proportional to their norm and the rate $\gamma$ on top of a deterministic nonunitary dynamics determined by the Hamiltonian in Eq.~\eqref{eq:Heff}. 

Notice that due to the specific form of the jump operators $G^\dagger_n G_n = 1$, the non-Hermitian contribution to the Hamiltonian reduces to a trivial decrease of the state norm $\propto N \delta t \gamma$.
Their action in Eq.~\eqref{eq:jump_app}, instead, acts trivially only if $\ket{\psi(t)}$ is an eigenstate of $G_n$; thus, when the jump operators are repeatedly applied to $\ket{\psi(t)}$, they induce dephasing between states belonging to different gauge sectors. This, in turn, suppresses the speed of the population transfer between them caused by the action of $H_{\rm err}$. 

\begin{figure}
    \centering
    \includegraphics[width=\columnwidth]{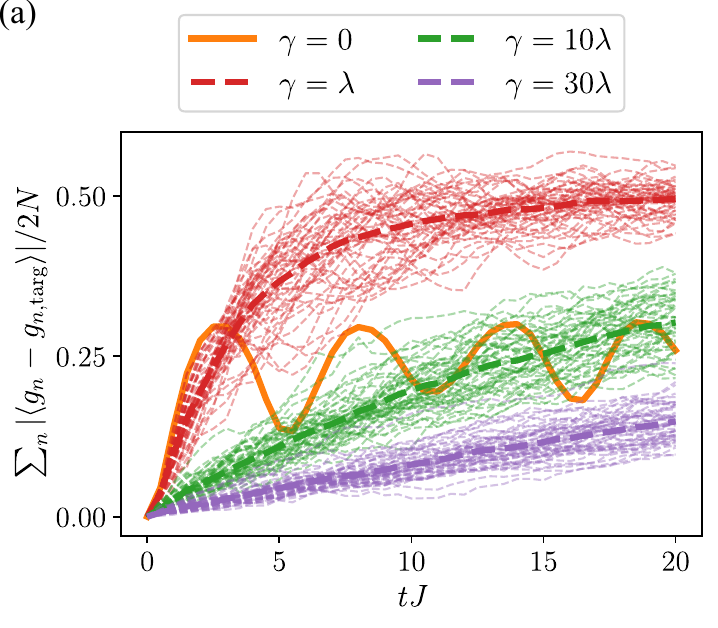}
    \includegraphics[width=\columnwidth]{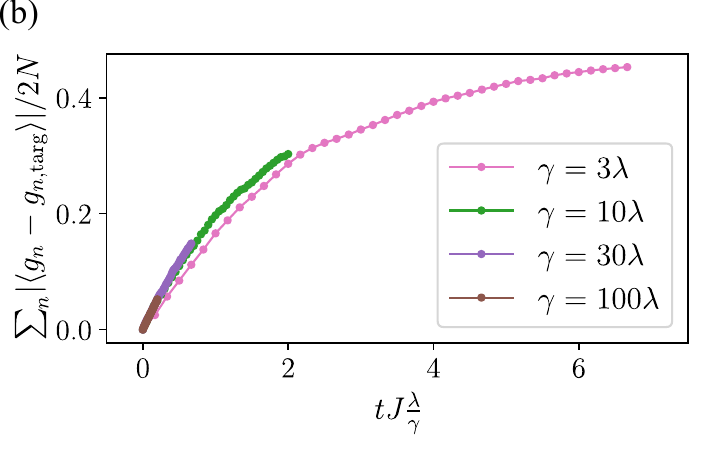}
    \caption{(a) Evolution of the average violation of the local charges, comparing the coherent error (solid orange line) and the corrected dynamics with different measurement rates (dashed lines). The thick lines represent the average over 50 different trajectories (fainter lines). Individual trajectories remain close to the ensemble average and the gauge violation decreases as $1/\gamma$ in the protected phase ($\gamma > \lambda$).
    The initial state corresponds to a local charge defect at the center of the chain and the physical parameters are $f=0.5J$, $\mu=0$, and $\lambda=0.3J$.
    (b) By plotting the average gauge violation versus the rescaled time $t\lambda/\gamma$, a perfect data collapse is achieved once the system is sufficiently deep in the quantum Zeno regime.
    }
    \label{fig:continuous_deconfined}
\end{figure}

We now focus on the dynamics of individual quantum trajectories, which has not been analyzed previously in the context of dissipative gauge protection. To this end, we consider a chain of $N=9$ matter sites with OBC ($L=17$ spins in total) prepared with a single matter defect in the central site $n=4$. 
The system is evolved with the Hamiltonian in the deconfined phase ($f=0.5J$, $\mu=0$) and with a coherent error strength $\lambda=0.3 J$.
To investigate the error protection effectiveness, we analyze in Fig.~\ref{fig:continuous_deconfined} the evolution of the gauge violation $\delta g$ [defined in Eq.~\eqref{eq:gauge_v}].
We show both the deterministic dynamics affected by the coherent error and the result of the continuous measurement protocol. For the latter, the average over 50 trajectories is plotted as a thick dashed line while individual stochastic dynamics are the fainter ones.

Large measurement rates provide an effective error suppression, although a drift outside the target sector deteriorates the quality of the simulation, requiring a higher measurement rate for reliably reaching a longer total evolution time. The trajectory spread also increases in time, although the data for the weakest rate ($\gamma=\lambda$) suggest that it saturates as soon as equilibrium is approached.
Since the steady state of Eq.~\eqref{eq:app_Lindblad} is the infinite-temperature density matrix, irrespective of the rate $\gamma$, equilibrium is always characterized by an average gauge violation of $1/2$. 

Despite the coherent error of $\lambda=0.3J$ being rather strong, the dynamics of the gauge violation is entirely dominated by the jump processes.
Indeed, if the coherent error alone is active, without gauge protection, a fast gauge violation is generated that then oscillates around an average value dependent on $\lambda$. 
In contrast, gauge protection via measuring the local charges causes a monotonous increase in the average violation. 
Only at the level of individual trajectories, do some oscillations survive but without any resemblance to the coherent dynamics.
In the Zeno regime ($\gamma > \lambda$), the steady increase of the gauge violation happens on a time scale set by the measurement rate $\gamma$, consistently with the behavior of the Liouvillian eigenvalues shown in Fig.~\ref{fig:liouv_es}.
Indeed, if we rescale time as $t\to t \frac{\lambda}{\gamma}$, the curves describing the average gauge violation for different $\gamma$ in time perfectly collapse on top of each other, as we report in panel (b) of Fig.~\ref{fig:continuous_deconfined}. A slight violation can be discerned only for the smallest $\gamma$ considered. 
When $\gamma<\lambda$, instead, the relaxation rate increases with $\gamma$, as seen from the Liouvillian eigenvalues, leading to a fast equilibration to an infinite temperature state characterized by $\langle \delta g \rangle =0.5$.

\section{Effective Liouvillian spectrum in $\mathbb{Z}_3$ and $U(1)$ gauge groups.} \label{app:Z3_and_U1}

Our analysis is not restricted to the simple $\Ztwo$ model discussed in the main text. In this section, we extend the investigation of the Liouvillian spectrum associated with the continuous-time limit of the DPS protection protocol to LGTs in one spatial dimension with $\mathbb{Z}_3$ and $U(1)$ symmetries. 
For both, the structure of the Hamiltonian is the same as Eq.~\eqref{eq:H} but the operators acting on the gauge degrees of freedom and the local charges need to be changed to take into account the larger groups.

\subsubsection{$\mathbb{Z}_3$ group}

A LGT with discrete $\mathbb{Z}_N$ gauge group is a generalization of the simpler model analyzed in the main text.
We consider generalized Pauli matrices (clock operators) satisfying the algebra
\begin{subequations}
    \begin{align}
        & Q^N = P^N = \mathbb{1} \ ,\\
        & QP = PQ \nep^{i \frac{2\pi}{N}} \ ,\\
        & Q |m\rangle = \nep^{i\frac{2\pi} {N}m}\ket{m}\ .
    \end{align}
\end{subequations}

Here, we focus on $N=3$.
After a Jordan--Wigner transformation on the fermions (matter sites), the Hamiltonian \eqref{eq:H} then becomes
\begin{eqnarray}
    H_{\mathbb{Z}_3} = & J\sum_n \left( \sigma^+_n P_{n,n+1} \sigma^-_{n+1} + {\rm h.c} \right)   \\
+ & \frac{f}{2} \sum_n (Q_{n,n+1}+Q^\dagger_{n,n+1})  + \frac{\mu}{2} \sum_n (-1)^n \sigma^z_n \nonumber  \ .
\end{eqnarray}
The gauge transformations are~\cite{Burrello2015} 
\begin{eqnarray}\label{eq:cloc_z3}
    G_n = Q_{n-1,n} Q^\dagger_{n,n+1}\nep^{-i\frac{\pi}{3}(1+\sigma^z_n)}
\end{eqnarray}
and their eigenvalues are the cubic roots of unity $g_n = \nep^{i\frac{2\pi}{3}m}$, with $m=0,1,2$.

\subsubsection{Truncated $U(1)$}

$U(1)$ is the gauge group of quantum electrodynamics~\cite{Wiese_AnnPhys2013}, making it one of the most popular targets for quantum simulations of LGT. Being a continuous group, we need to truncate it to map it on the discrete degrees of freedom of quantum hardware. 
Here, we adopt the quantum-link-model formulation for the gauge fields~\cite{Wiese_AnnPhys2013,Chandrasekharan_NucPhysB1997}.
In practice, we consider the operators $S^z$ and $U$ acting on the links, such that
\begin{eqnarray}
    S^z\ket{m} = m\ket{m} \ {\rm with} \ m= -S,-S+1, \dots, S \ ,
\end{eqnarray}
and 
\begin{equation}
    U\ket{m}=\ket{m+1} \ {\rm if}\ m<S \ .
\end{equation}
Using Kogut--Susskind staggered fermions, followed by a Jordan--Wigner transformation to hard-core bosons, the Hamiltonian of the theory is
\begin{eqnarray}
    H_{U(1)} = & J\sum_n \left( \sigma^+_n U_{n,n+1} \sigma^-_{n+1} + {\rm h.c} \right)   \\
+ & f \sum_n \left( S^z_{n,n+1}\right)^2  + \frac{\mu}{2} \sum_n (-1)^n \sigma^z_n \nonumber  \ ,
\end{eqnarray}
while the local Gauss' law generators read
\begin{equation}\label{eq:cloc_u1}
    G_n=S^z_{n,n+1} - S^z_{n-1,n} -\frac{1}{2}((-1)^n + \sigma^z_n) \ .
\end{equation}
The physical subspace is the one with $G_n=0$ and the $(-1)^n$ factor takes into account the different values of the charge on even and odd sites.
We truncate the gauge fields to $S=1$, meaning that the electric field can take three different values on each link.
The Hilbert space is, therefore, the same as the $\mathbb{Z}_3$ LGT but the local symmetries and the constraints they impose are different.

\begin{figure*}
    \centering
    \includegraphics[width=\linewidth]{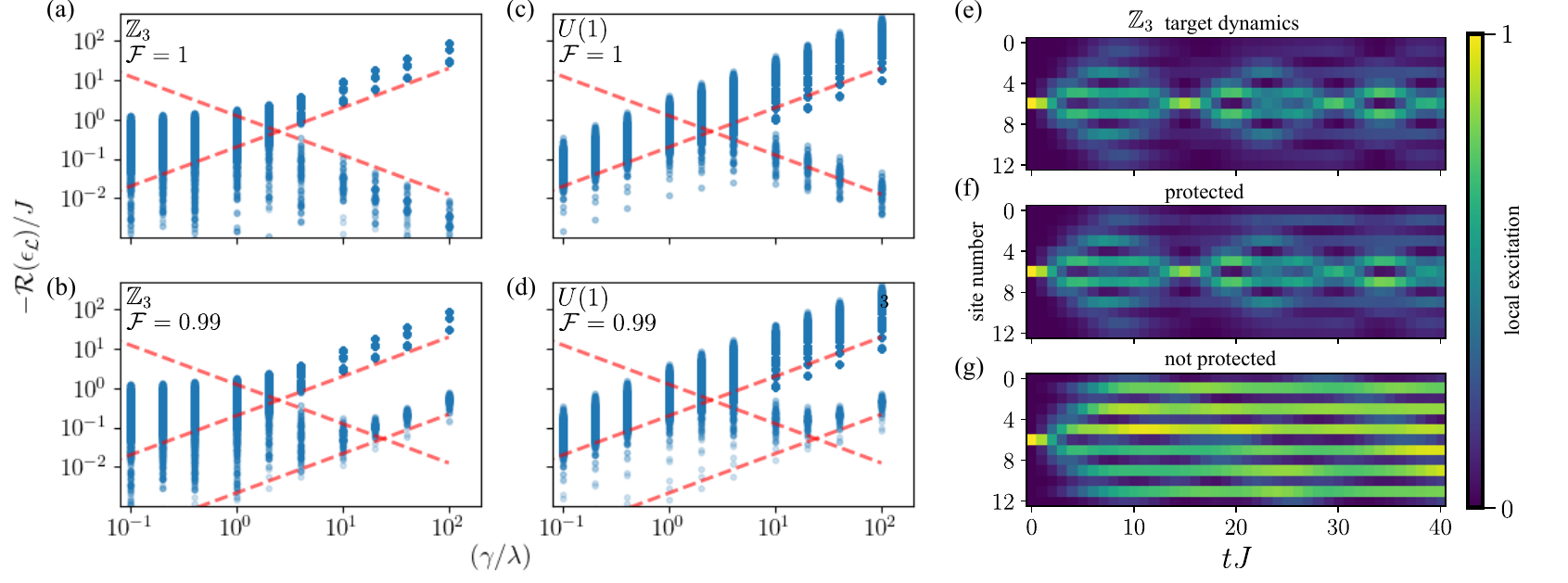}
    \caption{(a-d): Liouvillian spectrum as a function of the measurement frequency in the presence of unitary errors for the $\mathbb{Z}_3$ (a-b) and $U(1)$ (c-d) gauge groups. The branching in the spectrum signals the onset of the quantum Zeno phase that protects gauge invariance and survives also for finite measurement fidelity, see panels (b) and (d). The opacity of the markers reflects the density of states.
    (e-g): example of dynamics in the deconfined phase of a 1+1d $\mathbb{Z}_3$ LGT model. The Zeno protection with $\gamma = 20 \lambda $ (f) recovers the target dynamics (e), while it is almost completely lost (g) without it.
    Simulation parameters: $f=0.5J$, $\mu=0.1J$, $\lambda=0.2J$. In panels (a-d) the chain consists of $N=3$ matter sites, while in panels (e-f) $N=7$.}
    \label{fig:Z3}
\end{figure*}

\subsubsection{Liouvillian spectrum}

Following the gauge-symmetry-violating Hamiltonian used in the main text, here we consider two sources for coherent errors, rotations of the gauge fields and unassisted hopping between matter sites. The latter is independent of the gauge group and is identical to the one in the main text.
On-site electric field rotations take the form
$(P_{n,n+1}+P^\dagger_{n,n+1})$ for $\mathbb{Z}_3$ and $(U_{n,n+1}+U^\dagger_{n,n+1})$ for $U(1)$.

To investigate the presence of a quantum Zeno transition, we consider an effective Lindbladian where the unitary evolution is given by the model Hamiltonian and the coherent error terms with a strength $\lambda$, while the collapse operators correspond to the local charges in Eqs.~\eqref{eq:cloc_z3} and \eqref{eq:cloc_u1}.
Measuring the generators of gauge transformations for larger groups is technically more complicated than the simple stabilizer of the $\Ztwo$ group. A circuit implementation can be found, for instance, in Ref.~\cite{Stryker_PRA2019}.
We also consider the situation in which the measurements of the local charges have finite fidelity, by adding further on-site dissipative noise with a rate $\gamma (1-\mathcal{F})$, where $\gamma$ is the measurement frequency and $\mathcal{F}$ their average fidelity.

Figure~\ref{fig:Z3} summarizes our results: the appearance of the quantum Zeno phase in the Liouvillian spectrum, see panels (a-d) is qualitatively similar to what we observed in Sec.~\ref{sec:dps} for both $\mathbb{Z}_3$ and $U(1)$ groups, even though the critical measurement frequency slightly changes. The Zeno phase survives also with finite measurement fidelity (here $\mathcal{F}=0.99$) on a finite range of $\gamma$, as observed for the $\Ztwo$ LGT model.
When looking at a specific example of gauge-invariant dynamics one wishes to recover, the effect of the measurements is again similar to what we observed for the $\Ztwo$ model. 
We report an instance for the $\mathbb{Z}_3$ group in panels (e-g) of Fig.~\ref{fig:Z3}, where a dramatic improvement in the Zeno phase becomes clear with respect to the bare evolution subject to coherent errors.

\section{Trajectory fluctuations in analog quantum simulations}
\begin{figure}
    \centering
    \includegraphics[width=\columnwidth]{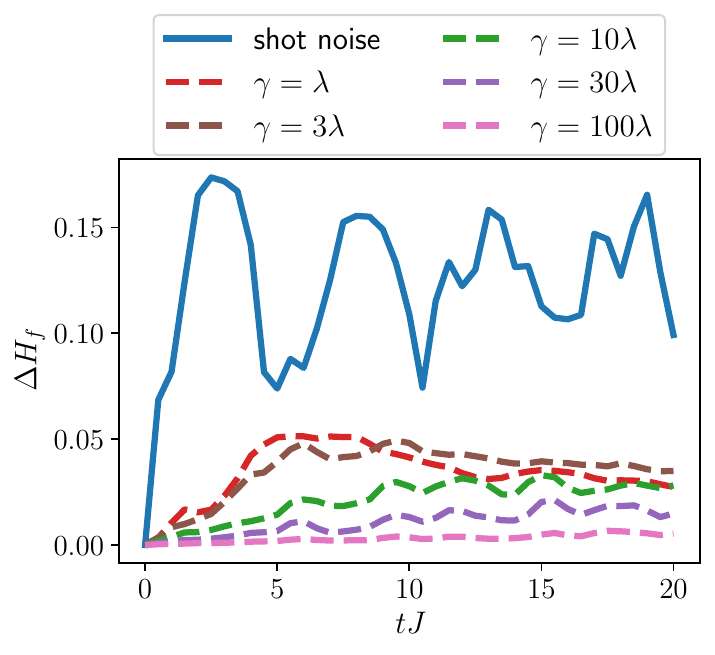}
    \caption{Evolution of the standard deviation of the electric field. The quantum variance measured on the ideal evolution (solid line) dominates over the fluctuations due to the quantum trajectory stochastic dynamics (dashed lines).}
    \label{fig:fluctuations}
\end{figure}

One important issue is whether the fluctuations due to the stochastic dynamics induced by the quantum measurement are larger than the quantum variance of the operator we wish to measure. 
In other words, does this protection mechanism introduce an overhead in the number of runs needed to evaluate the expectation values of observables of interest?

To answer this question, we compare in Fig.~\ref{fig:fluctuations} the time evolution of the standard deviation of the electric field $\Delta H_f $ due to the stochastic trajectories (dashed lines) and its shot noise $\langle H_f^2 \rangle - \langle H_f \rangle^2 $ computed on the gauge-invariant physical state evolved with $H_0$ (solid line).
Even for a weak measurement rate, where measuring the local charges is detrimental for gauge protection, the quantum variance of the electric field dominates over that of the quantum trajectories, which instead becomes progressively smaller as $\gamma$ increases.
This analysis holds when the system is not in an eigenstate of the observable of interest, since in that case the shot noise vanishes while the same does not happen to the fluctuations of the stochastic trajectories.
However, we expect the shot noise to be larger in general out-of-equilibrium states that constitute the relevant scenario for the time evolution after a quantum quench.
Thus, in general, the number of measurements required to estimate the average electric field does not significantly increase as a result of the continuous measurements of the local gauge symmetries.

\section{Trajectory post-processing}\label{app:filter}
In Sec.~\ref{ssec:ec}, we adopted an outlier exclusion method to automatically filter out trajectories in which the error correction failed because of non-detectable errors affecting neighboring symmetry generators.
Here we give more details on the method used for the data in in Fig.~\ref{fig:dynamics_full_error} (dashed green line and triangular markers).
At each Trotter step, we check the distance between an individual trajectory and the average in units of the standard deviation of the observable. 
If at any point this distance is larger than a certain threshold, we discard the trajectory, assuming that a non-detectable error occurred. 
This process can be iterated several times to further refine the ensemble of trajectories used to compute the corrected dynamics. 

\begin{figure}
    \centering
    \includegraphics[width=\columnwidth]{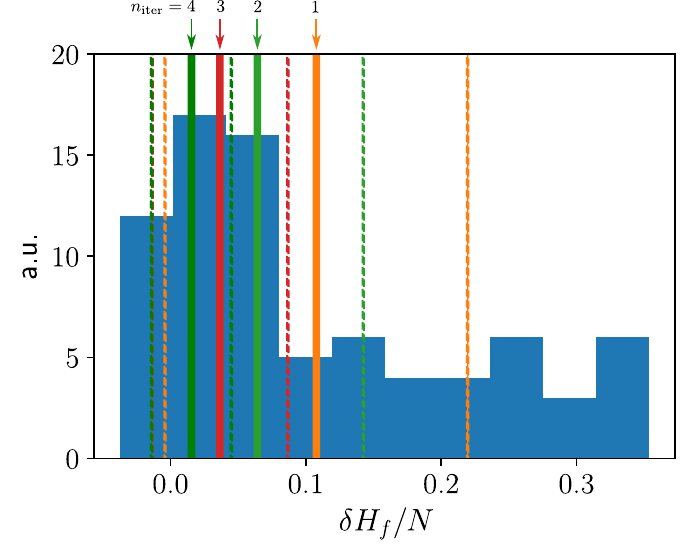}
    \caption{Example of the improvement by post-selecting the trajectories. The histogram is the distribution of $\delta H_f/N$ at $t J=10$ over 100 trajectories, while the vertical lines indicate the average (solid lines) and the standard deviation (dashed lines) obtained after each iteration of the filtering process. }
    \label{fig:filter}
\end{figure}

We report in Fig.~\ref{fig:filter} the result of such a method for an error probability $p_{\rm err}=0.02$ and a coherent error strength $\lambda=0.2J$, at a snapshot $tJ=10$ with $n_t=20$ Trotter steps.
We focus again on the difference $\delta H_f$ between the expectation value of the electric field Hamiltonian obtained with the DPS gauge protection and the ideal evolution.
On top of the distribution of $\delta H_f$ extracted from 100 stochastic trajectories, we plot the average (solid lines) and the corresponding standard deviation (dashed lines) after several rounds of post-selection.
At each iteration, we discard all trajectories that at any point in the evolution were farther than $2\sigma$ from the ensemble average.  
This procedure improves considerably the estimate of the ideal dynamics of $\langle H_f \rangle$, as well as reducing its uncertainty due to the trajectory dispersion.
Figure~\ref{fig:filter} clearly shows that the post-selection procedure automatically identifies the bundle of trajectories that follows more closely the exact time evolution while it discards those where undetected errors generate unphysical results without breaking the gauge invariance.

As the shot noise of $H_f$ remains quite large, see Fig.~\ref{fig:fluctuations} for comparison, it is hard to predict the efficiency of trajectory post-selection in a realistic experiment.
It is possible, however, that machine-learning clustering methods will still be able to discriminate between measurements where undetected errors occurred and those where the correction was successful.

%

\end{document}